%% file: apssamp.tex
\documentclass[aps,physrev,reprint,groupedaddress]{revtex4-2}
\usepackage{xcolor}
\usepackage{graphicx}
\usepackage{hyperref}
\usepackage{tabularx}
\usepackage{xtab,afterpage,longtable}
\begin{document}

%\preprint{}

\title{Evolution of robust cell differentiation under epigenetic feedback}


\author{Davey Plugers}
\email{davey.plugers@nbi.ku.dk}
%\thanks{}

\author{Kunihiko Kaneko}
\email{kunihiko.kaneko@nbi.ku.dk}
\affiliation{Niels Bohr Institute, University of Copenhagen, Blegdamsvej 17, 2100 Copenhagen, Denmark}

%\paragraph{}%\collaboration{}} 
\date{\today}

\begin{abstract}
In multi-cellular organisms, cells differentiate into multiple types as they divide. States of these cell types, as well as their numbers, are known to be robust to external perturbations; as conceptualized by Waddington’s epigenetic landscape where cells embed themselves in valleys corresponding to final cell types. How is such robustness achieved by developmental dynamics and evolution? To address this question, we consider a model of cells with gene expression dynamics and epigenetic feedback, governed by a gene regulation network. By evolving the network to achieve more cell types, we identified three major differentiation processes exhibiting different properties regarding their variance, attractors, stability, and robustness. The first of these, type A, exhibits chaos and long-lived oscillatory dynamics that slowly transition until reaching a steady state. The second, type B, follows a channeled annealing process where the epigenetic changes in combination with noise shift the cells towards varying final cell states that increase the stability. Lastly, type C exhibits a quenching process where cell fate is quickly decided by falling into pre-existing fixed points while cell trajectories are separated through periodic attractors or saddle points. We find types A and B to correspond well with Waddington’s landscape while being robust. Finally, the dynamics of type B demonstrate a differentiation process that uses a directed shifting of fixed points, visualized through the dimensional reduction of gene-expression states.  Correspondence with the experimental data of gene expression variance through differentiation is also discussed.
%In multicellular organisms, cells differentiate into multiple types as they divide. This process, along with the developmental trajectory, is remarkably robust to external perturbations, a concept captured by Waddington’s epigenetic landscape, where cells settle into valleys corresponding to final cell types. How is such robustness in developmental pathways, known as homeorhesis, ensured by developmental dynamics and evolution? To address this question, we investigate a model of cells with gene expression dynamics and epigenetic feedback, governed by a gene regulatory network. By evolving the network to generate a greater diversity of cell types, we identify three major differentiation mechanisms, each exhibiting distinct properties in terms of variance, attractors, stability, and robustness. In type A, differentiation follows chaotic and long-lived oscillatory dynamics, gradually stabilizing into a steady state. In type B, epigenetic changes and noise facilitate a channeled annealing process, dynamically shifting the cellular landscape toward varying final states. In type C, cell fate is rapidly determined through a quenching process, where trajectories fall into pre-existing fixed points, separated by periodic attractors or saddle points. We find that types A and B align well with Waddington’s landscape and exhibit high robustness to noise and initial condition variations. Notably, type B dynamics reveal a novel mechanism for achieving homeorhesis during differentiation.
\end{abstract}

\keywords{Multicellular Differentiation, Epigenetic Feedback, Evolved Gene Regulatory Networks, Differentiation Type Classification, Robustness}


\maketitle

%\subsection{}
%\subsubsection{}

\section{Introduction}
\input{Introduction}

%%%
\section{Model}
\input{2.TD/TechDet_0Intro}
\subsection{Gene-Expression-Epigenetic-Modification Model}
\input{2.TD/TechDet_0Model}
\vspace{-10mm}
\subsection{Evolution}
\vspace{-3mm}
\input{2.TD/TechDet_1Evo}
\vspace{-8mm}
\subsubsection*{Initialisation}
\vspace{-5mm}
\input{2.TD/TechDet_2Evo}
\vspace{-5mm}
\subsubsection*{Dynamics}
\vspace{-5mm}
\input{2.TD/TechDet_3Evo}
\vspace{-13mm}
\subsubsection*{Fitness, Selection and Mutation}
\vspace{-5mm}
\input{2.TD/TechDet_4Evo}

%%%
\vspace{-5mm}
\section{Three Typical Differentiation Processes}
\input{3.TTM/Mechanisms_0Intro}
\vspace{-3mm}
\subsection*{Classification through initial attractor}
\input{3.TTM/Mechanisms_Classification2}
\vspace{-7mm}
\subsection{Type A: Oscillation-fixation}
\input{3.TTM/Mechanisms_A}
\subsection{Type B: Channelled Annealing}
\input{3.TTM/Mechanisms_B}
\subsection{Type C: Quenching}
\input{3.TTM/Mechanisms_C}
\subsection*{Noise Level Dependence}
\input{3.TTM/Mechanisms_Statisitcs}
\vspace{-20mm}
\subsection*{$\nu$-dependence}
\input{3.TTM/Mechanisms_nu}

%%%
\section{Analysis}
\vspace{-5mm}
\input{4.Analysis/Analysis_0Intro}
\vspace{-5mm}
\subsection{Statistics of attractor types with the change in $\theta$}\label{sec:Attractor}
\input{4.Analysis/Analysis_1Attr}
\vspace{-5mm}
\subsection{Inter- and Intra-cellular Variance}\label{subsec:Var}
\input{4.Analysis/Analysis_2Var}
\subsection{Robustness}
\input{4.Analysis/Analysis_3Robust}

%%%
\section{Discussion}
\input{5.Disc/0MainDisc}
%\subsection{Experimental Relevance}
\input{5.Disc/1ExpRel}

%%%%
\vspace{-5mm}
\begin{acknowledgments}
 We are supported by the Novo Nordisk Foundation Grant No. NNF21OC0065542.
We would also like to thank Yuuki Matsushita, Namiko Mitarai, Kim Sneppen, Riz Noronha, and Stanley Brown for the illuminating discussions. The data that support the findings of this article are openly available \footnote{\href{https://github.com/DaveyPlugers/Evolution_Of_Robust_Cell_Differentiation_Under_Epigenetic_Feedback}{Public GitHub Repository by Davey Plugers: https://github.com/DaveyPlugers/Evolution\_Of\_Robust\_\\
Cell\_Differentiation\_Under\_Epigenetic\_Feedback}   \\ Animations of several figures, as well as networks and scripts to analyze results have been made available. Created on 27/03/2025 \label{fn:Git}}.
\end{acknowledgments}

%%%%%
\vspace{-5mm}
\appendix*
\section{Supplementary figures}
\input{Appendix.tex}
\bibliography{apssamp.bib}

\end{document}

%% file: Introduction.tex
Epigenetics was introduced by Conrad Waddington in 1942 to explain the generation of differentiated cell types. He described this phenomenon as a mechanism in development that bridges the gap between genes and phenotype\cite{waddington1942epigenotype}, proposing an epigenetic landscape\cite{waddington1957strategy} as a visual metaphor for cellular differentiation, where cells branch into various developmental pathways. As cell fate commitment progresses, the branching of valleys represents differentiation into distinct pathways making certain cell fates irreversibly inaccessible, leading to the loss of pluripotency. The expression or repression of genes controls the underlying shape of the landscape (valleys and hills), and Waddington identified this gene expression control as the epigenetic mechanism\cite{waddington1942epigenotype}.\\
\\
The study of epigenetics remained relatively inactive until the 1990s, during which novel research on chromatin structure modifications revealed the connection between methylation and acetylation of histones with the expression levels of genes\cite{deichmann2016epigenetics,willbanks2016evolution,stillman2018histone}. These chemical modifications are considered major mechanisms within epigenetic regulation and play a crucial role during development\cite{atlasi2017interplay}. The change in chromatin structure adjusts accessibility for transcription and determines gene expression feasibility over longer time scales than the change in protein expression levels\cite{zhang2019metabolic}. \\
\\
In short, the epigenetic modification process can be explained as follows: Gene expression leads to the synthesis of mRNA, which generates the associated proteins\cite{jacob1961genetic}. These proteins can regulate a promoter, thereby affecting the chromatin structure of the genes\cite{atlasi2017interplay,kiefer2007epigenetics,gibney2010epigenetics,rando2009genome,hihara2012local}. The resulting structural changes either suppress or promote mRNA synthesis, depending on the protein-promoter interaction\cite{kiefer2007epigenetics,weinhold2006epigenetics}. Eventually, these effects lead to a stabilized cellular state, where epigenetic fixation ensures robustness in gene expression patterns and intracellular protein expression levels\cite{grunstein1998yeast,gibney2010epigenetics,reik2007stability,cortini2016physics,rue2015cell}. Theoretically, such gene expression dynamics leading to different cellular states have been pioneered by Kauffman\cite{kauffman1969metabolic} and were investigated extensively. Dynamical systems studies have elucidated the roles of oscillations\cite{Furusawa2012}, transition states\cite{brackston2018transition}, geometry, basins of attraction, and potential landscapes \cite{jaeger2014bioattractors,wang2010potential,rand2021geometry} in achieving differentiation. These models exhibit the major features of differentiation\cite{villani2011dynamical} but have not yet fully explained the evolutionary process of hundreds of attractor states representing multiple cell fates\cite{newman2020cell}. Slow epigenetic fixation is modeled through dynamical systems theory by considering the (protein) expression levels of genes and their epigenetic modification levels\cite{karlebach2008modelling}. Gene expression dynamics describe how proteins regulate gene promoters, controlling the activation or repression of specific genes. Meanwhile, slower epigenetic modifications lead to long-term chromatin changes which remain across multiple gene expression cycles\cite{sasai2013time}. This modification stabilizes gene expression patterns and in the end, fixates the cellular state\cite{azuara2006chromatin,ng2008epigenetic}. Understanding how dynamical systems with fast gene expression changes and slower epigenetic modifications lead to robust cell differentiation remains a fundamental research goal.\\
\\
Such dynamical systems models with gene expression and epigenetic modifications have been studied previously. These studies successfully generate fixed-point attractors that represent distinct final cell states\cite{Matsushita_2020, miyamoto2015pluripotency} and are capable of recreating cellular reprogramming\cite{matsushita2022dynamical}. By assuming oscillatory gene expression dynamics\cite{Furusawa2012}, these models successfully reproduced Waddington’s landscape, capturing hierarchical branching, homeorhesis, and robustness. However, these prior studies were limited by their reliance on specific network configurations to induce oscillations and their dependence on initial conditions to generate distinct final cell states. In these previous studies, networks were randomly selected, and only those exhibiting differentiation into multiple stable states (resulting from oscillatory behavior and adjustments to the initial conditions) were analyzed. Although the relevance of oscillatory expression was noted in these studies, such oscillations were not consistently observed in experimental data. Furthermore, alternative differentiation processes were not explored. Additionally, the role of stochasticity, an essential factor in gene expression, was not addressed.\\
\\
Here, we evolve gene regulatory networks (GRNs) to investigate how the interplay between gene expression dynamics and epigenetic modifications drives cell differentiation, as envisioned in Waddington’s epigenetic landscape. Starting from single cells with identical initial conditions, we examine how stochastic noise in gene expression levels enables differentiation into multiple cell fates. Networks capable of generating multiple cell types, enabled by the stochasticity breaking homogeneity, are selected through an evolutionary algorithm. The selected GRNs promote robust multicellular development, resulting in diverse final cell fates. Through a dynamical systems analysis, we identified three fundamental differentiation processes.\\
\\
Type A networks (Oscillation-Fixation): Gene expressions initially exhibit chaotic dynamics, characterized by a positive Lyapunov exponent. These dynamics gradually stabilize due to epigenetic modifications. Epigenetic fixation suppresses the chaotic stage and leads cells toward fixed point attractors at later times. The initial chaotic attractor contributes to robust differentiation in response to perturbations in initial conditions. \\
\\
Type B networks (Channelled Annealing) : Gene expressions initially exhibit weakly stable fixed points, and cell differentiation occurs  through the migration of fixed-point attractors. Their fixed point shifts randomly due to noise, while epigenetic modifications introduce directional changes in gene expression, enhancing stability. \\
\\
Type C networks (Quenching): Gene expressions initially exhibit stable fixed points with on/off states or limit cycles. Multiple fixed points emerge within a short time span.\\
\\
We analyzed these three differentiation processes by examining gene expression level activity, maximal Lyapunov exponents, variance in gene expression levels, and the robustness of final cellular states to perturbations. Finally, we evaluate whether each of the three differentiation types reconstructs Waddington’s landscape and compare our findings with experimental data, emphasizing the significance of type B.

%% file: 2.TD/TechDet_0Intro.tex
This work can be divided into three simulation procedures that build upon each other. To start, we have the model consisting of a gene regulatory network with epigenetic feedback, which directs the gene expression dynamics within a cell. Secondly, we have an evolutionary algorithm: by evaluating the final gene expression levels and computing fitness, we select the networks with higher fitness among the mutated variants. Finally, using the evolved networks, we simulate the intracellular gene expression level dynamics while measuring and sampling their properties. In this section, we describe the model we apply and the evolutionary algorithm used to obtain these networks.

%% file: 2.TD/TechDet_0Model.tex
We use a gene regulatory network with slow epigenetic modifications to model intracellular dynamics. Here, we adopt the model by\cite{mjolsness1991connectionist} (see also\cite{salazar2001phenotypic,kaneko2007evolution}). This model is represented by the following stochastic differential equation \ref{eq:Main} for the gene expression level $x_i(t)$, with the coupled slow-timescale equation \ref{eq:Epi} for the epigenetic fixation $\theta_i(t)$.\\
\vspace{-2mm}
\begin{equation} \label{eq:Main}
    \frac{dx_i}{dt} = \left[F\left(\sum^{M}_j \frac{J_{ij}}{\sqrt{M}}x_j + \theta_i + c_i \right) - x_i\right] + \sigma \eta\textcolor{red}{_i}
\end{equation}
\begin{equation} \label{eq:Epi}
    \frac{d \theta_i}{dt} = \nu(x_i - \theta_i)
\end{equation}\\
Gene expression levels $x_i$ (for $i \in \{1,...,M=40\}$) are used to represent protein concentration, which promotes or suppresses the expression of other genes. This gene-interaction is modeled through a gene regulatory matrix (GRN) whose matrix elements are denoted by $J_{ij}\in\{-1,0,1\}$, where -1, 0, 1 represent suppression, no interaction, and promotion, from other proteins $x_j$ onto $x_i$ respectively.\\
\\
The gene expression is modeled through a smooth step function $F(z) = tanh(\beta z)$, with steepness $\beta=40$ to create an on-off transition. Lower values for $\beta$ allow for more intermediate gene expression values; however, for lower values (e.g. less than 10), the stepwise behavior of $tanh(\beta z)$ is lost, and fitness is not easily increased as differentiation to on/off states is harder. Fully expressed state $(x_i = 1)$ follows if the gene interaction term $\sum^{M}_j \frac{J_{ij}}{\sqrt{M}}x_j$ exceeds the threshold -($\theta_i + c_i$); otherwise the expression approaches the unexpressed state $(x_i = -1)$. This activation threshold consists of a constant activation shift $c_i\in[-0.5,0.5]$ and the slow epigenetic fixation term $\theta_i$ from equation \ref{eq:Epi}.\\
\\
The $-x_i$ ensures our system is stabilized within the region $x_i\in[-1,1]$ as the image of the activator function $F(z)$ is restricted to this codomain. Finally, we introduce $\eta_i$, a Gaussian white noise process with variance 1, multiplied by the noise strength $\sigma$ to represent the stochastic effects. This last term, $\sigma$, represents noise in the expression levels breaking symmetry between cells and making the dynamics stochastic. This accounts for intracellular noise in the gene expression, known as stochastic gene expression\cite{raj2008nature,elowitz2002stochastic,mcadams1997stochastic,vinuelas2013quantifying,kaern2005stochasticity,furusawa2005ubiquity}. While the relevance of multiplicative noise to stochastic gene expression dynamics has been discussed \cite{Coomer2022}, here we adopt the standard additive noise as it can be analyzed easily. In the present study, the role of noise is to perturb attractors or to shift fixed points. As our simple model already exhibits these properties, there is no qualitative change between additive and multiplicative noise. The three types that will be discussed exist independently of the choice of noise.\\
\\
Next, following\cite{Matsushita_2020}, epigenetic modification is introduced by Equation \ref{eq:Epi}, which changes the feasibility of expression so that the threshold level $-(\theta_i+c_i)$ changes, i.e., $\theta_i(t)$ represents the dynamically changing modification level. This modification level changes with time depending on the expression level. Generally, when a gene is expressed (not expressed), the modification occurs so that the feasibility of expression is increased (decreased), respectively. Hence, there is a positive feedback loop between the expression level and modification\cite{dodd2007theoretical,sneppen2008ultrasensitive}. In summary, the expression levels $x_i$ increase or decrease based on the sign of the activator function's argument. Eventually, due to the epigenetic modification, they settle into a stable fixed point with a value of $\pm1$, representing either the fully expressed or non-expressed state.\\
\\
Here, $\nu$ represents the speed of modification level relative to that of expression whose inverse gives the time scale of the epigenetic modification. In general, change in epigenetic modification occurs much slower than the expression level itself. In the first set of simulations, following\cite{Matsushita_2020}, we adopt  $\nu = 6\cdot10^{-4}$, whereas larger values of $\nu$ (e.g. 0.01) are tested, which show equal capacity of differentiation. As $x_i$ approaches $\pm 1$, the corresponding modification level $\theta_i$ converges to a fixed point at ±1.\\
\\
The gene expression dynamics for fixed $\theta_i$ have been studied extensively\cite{salazar2001phenotypic,kaneko2007evolution,schlitt2007current,mjolsness1991connectionist}. Depending on the matrix $J_{ij}$, the system’s attractor can be chaotic or (quasi-) periodic, where gene expressions continuously switch on and off. In other cases, stable fixed points are reached, where gene expression remains constant. By including the change in the modification ($\theta_i$), the thresholds are gradually adjusted to match the gene expression levels, causing the expression levels to be drawn toward fixed points. Eventually, the time-dependent attractors are replaced by stable fixed points, with gene expression levels $x_i$ and epigenetic modification levels $\theta_i$ converging to $\pm1$.\\
\\
Due to the presence of a stochastic noise term, the system is no longer fully deterministic, thus requiring a statistical approach. The different stochastic perturbations experienced by individual cells makes it so that each cell may converge to different fixed points. Understanding the distribution of final cellular states is a crucial question for multicellular organisms. In our model, we address this by initializing multiple cells with the same GRN and initial conditions. Through cell divisions, we generate multiple cells and analyze the distribution of their final fixed points. Specifically, we perform six successive divisions, yielding  $N=2^6=64$ individual cells, and then explore the gene expression dynamics of each cell. Here, we neglect cell-cell interactions and external environmental inputs. The final distribution of cell states acts as a model for how a multicellular organism might look after differentiation. Since cell-cell interactions are absent, a key advantage is that cellular dynamics can be studied independently, where all observed properties are attributed to the epigenetic modifications and gene expression level dynamics from the GRN.\\
\begin{figure}[!h]
    \centering \includegraphics[width=0.9\linewidth]{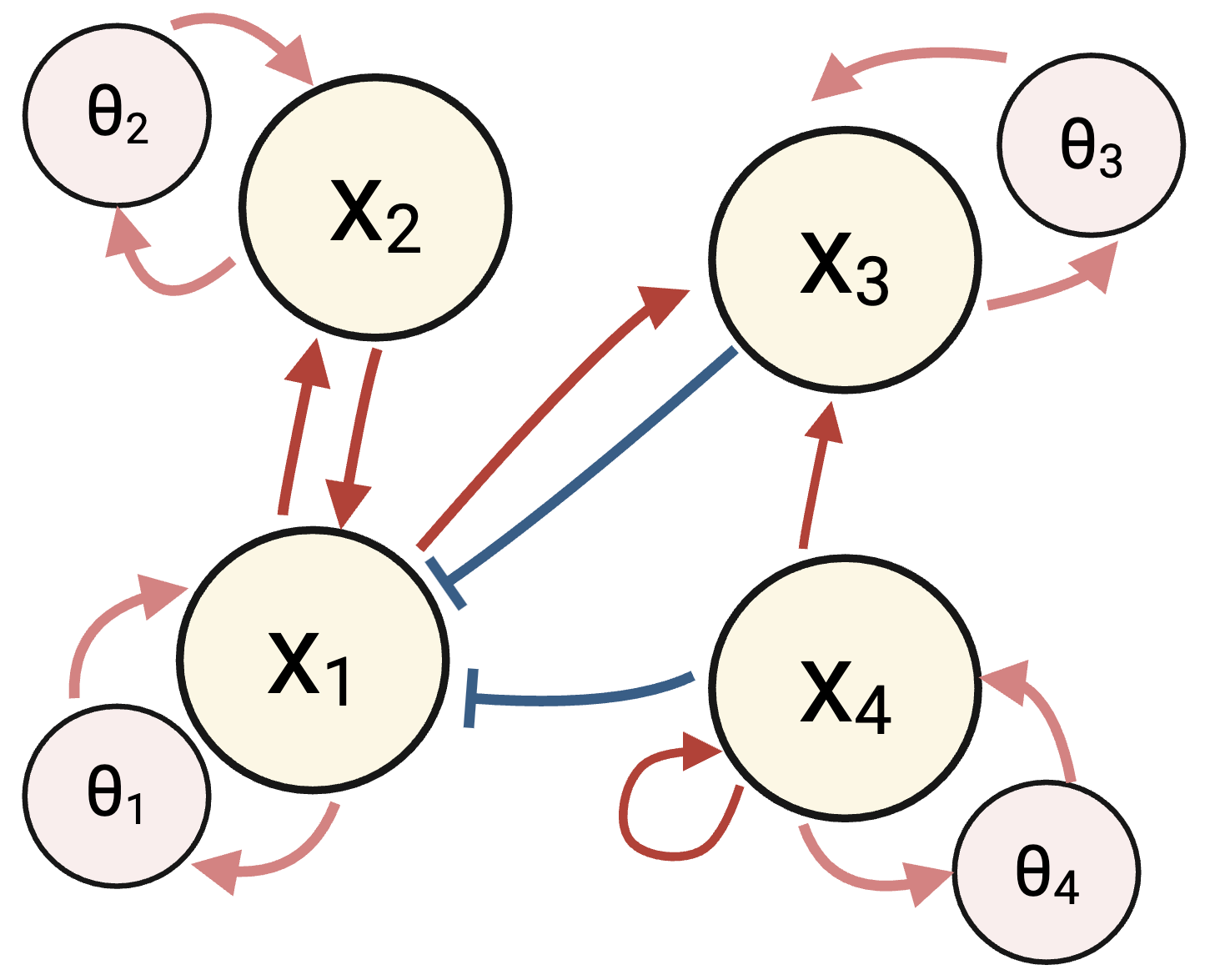}
    \caption{Schematic representation of a simplified network with M=4. All genes ($x_i$) have a positive feedback cycle with their epigenetic factor ($\theta_i$). Gene interactions are not symmetric and can promote (red) and suppress (blue) all genes, including themselves.}
    \label{fig:enter-label}
\end{figure}\\

%% file: 2.TD/TechDet_1Evo.tex
In this model, different cell types are given by distinct gene expression patterns represented by $x_i=\pm1$. Here, we postulate that organisms in the model have a higher fitness if more cellular types are achieved. Cells undergo dynamics depending on the GRN interactions by $J_{ij}$, these dynamics have to be optimized to achieve a greater number of cell types.
Therefore, we select networks that produce more distinct cell types by mutating the $J_{ij}$ connections. A more detailed description of this selection is as follows:

%% file: 2.TD/TechDet_2Evo.tex
In our setup, we keep initial gene positions $x_i(t=0)$, $c_i$ and the noise strength $\sigma$ fixed throughout the evolution procedure. For the noise, we choose $\sigma \in \{0.01, 0.04, 0.1 \}$, corresponding to low, medium, and high noise levels. During evolution, only the GRN matrix connections $J_{ij}$ vary as they are mutated across generations. We start with 32 different networks and run four identical copies of each, totaling 128 runs per generation. The GRN is generated using a directed Erdős–Rényi model with a connection probability of $p = 5/M$. Any non-zero connection $J_{ij}$ is assigned -1 or 1 with equal probability.  

%% file: 2.TD/TechDet_3Evo.tex
Each of the 128 runs consists of $N=64$ independent cells with M=40 genes $x_i$ and epigenetic factors $\theta_i$. The cells for a given network start from the same initial conditions but diverge due to the stochastic term $\sigma dW$, potentially leading to different cell fates. All cells are run until $t=20000$, which is 12 times longer than the slowest timescale $\tau_\theta = 1/\nu \approx 1666$ to ensure cells are fixated. Cell fate commitment is represented based on the gene expressions $x_i$ at the end of the simulation.\\
\\

%% file: 2.TD/TechDet_4Evo.tex
\begin{figure}[!h]
    \centering
    \includegraphics[width=0.84\linewidth]{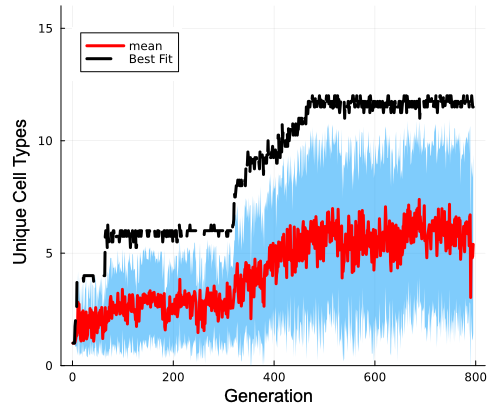}
    \caption{The fitness, i.e. the distinct cell types (y-axis) plotted as a function of generation (x-axis). The black line shows the best-performing network, while the red with the blue ribbon shows the mean and standard deviation in scores. The fitness increases stepwise. It stays stable for 50-350 generations until mutation around generation 350 connects one of the target genes such that it is now able to differentiate into about twice as many types.}
    \label{fig:evolution}
\end{figure}

Network selection is based on fitness, defined as the number of distinct cell types present in the final 64 cells of each network. Cell types are identified based on the on/off expression of the $N_t$ target genes. Here we set $N_t=4$ so that $x_1,x_2,x_3,x_4$ act as the target genes, with a maximum of $(\Omega = 2^N_t$) cell types. This is to ensure that the network creates a proper differentiation process targeted to specific genes.\footnote{For some networks, there are variations in the non-target gene expressions within the same cell type. These slight variations are uncommon and are quickly mutated away, and therefore, we consider them as the same type.}  \\
\\
There are $2^{N_t}$
possible cell types, determined by the target gene expression patterns. We require $N=64\geq 2^{N_t}$ for the count of individual cells to produce all requested cell types. Furthermore, as the number of target genes increases, it is more difficult to achieve the requested cell types. If $N_t$ is closer to the number of genes $M$, the evolution to achieve postulated cell types is harder. The existence of non-target genes is relevant for achieving distinct cell types robustly. We adopt 4 target genes among M=40 genes, as in this case, the robust differentiation process into distinct cell types is achieved through evolution. Reducing target cell types, by decreasing $N_t$ to 3 or less, does not change the differentiation processes that will be presented in this work. While increasing $N_t=6$ such that $N=64 = 2^{N_t}$ makes it difficult for a network to achieve high fitness, as each cell needs to differentiate to a unique cell type without any form of cell-cell communication. Further increasing the target genes such that $N=64 \ll 2^{N_t}$ breaks down any targeted differentiation process, making the gene dynamics highly chaotic and causing cell types to be irreproducible.\\
\\
The network size $M=40$ was chosen because a sufficiently large network is required to make chaotic orbits or limit cycles common. Mutations and genetic drift are also less destructive for larger networks since many genes will not play a role in the differentiation process, which eases the evolutionary process. It is also small enough to make the simulation and evolutionary durations numerically feasible. Furthermore, we have performed evolutionary runs for $M\in[10,20,80]$ where we reconfirmed the presence of the three types discussed in this work.\\
\\
The score of a given network is the mean number of distinct cell types generated across its four copies. The top 8 of the 32 networks are selected for the next generation. From the selected network, we create one direct copy without any mutation and three mutated versions. To mutate the network, we select one random non-zero connection $J_{ij}$ and set it to 0; we also take one $J_{ij}=0$ and set it to either -1 or 1 with equal probability. This ensures that the network’s connection density remains unchanged, preventing novel behaviors from arising purely due to changes in network density rather than the rearrangement of connections. This process is repeated up to 2000 generations to obtain GRNs that generate a sufficient number of differentiated cell types. As shown in Fig. \ref{fig:evolution}, GRNs with more than 12 cell types are commonly evolved across all simulation runs. The primary goal of this study is to explore and analyze typical differentiation processes rather than the evolutionary process itself. Thus, the evolutionary procedure here is not necessarily biologically realistic.\\
\vspace{-3mm}
%\subsubsection*{Network Parameter Space}}
%\textcolor{red}{
%Here add a section discussing the validity and sampling of the networks we used. Discuss that we stuck to Sparse networks like real GRNs however that we may need to consider that they follow power-law distributions (https://doi.org/10.1038/s41598-019-39866-z). Discuss how we have 5*M total connections whose removal are all sampled multiple times during evolution. Also the M*M = 1600 total possible connections are sampled multiple times as we have 2000*32*3 tries in an evolutionary run. While we may not have sampled the entire phase space of networks as this is impossible. By randomly generating and evolving for sufficiently long times we expect to have performed a sufficiently well sampling. One criticism point may be to use degree distributed networks, different sparsity, allow for weights or to not keep the sparsity fixed. As these things do limit our phase space exploration.}

%% file: 3.TTM/Mechanisms_0Intro.tex
Once the networks evolved and exhibited multiple cell types, we analyzed their underlying dynamical systems to understand how they achieved robust differentiation. We identified three distinct processes that frequently show up in network behavior. Each process possesses unique properties that contribute to the efficiency of robust cell fate decisions. These three types are classified according to the type of initial attractors.\\
\\
%Type A relies on \textcolor{red}{chaotic} oscillations that gradually narrow into sub-cycles due to epigenetic regulation, eventually stabilizing into multiple fixed points corresponding to distinct cell types. In type B, orbits quickly settle into fixed points but are gradually shifted by noise and epigenetic changes, causing cells to drift into diverging pathways that lead to distinct final states. In type C, multiple fixed points emerge rapidly, and cells reach them by crossing saddle points or through abrupt transitions from periodic attractors. \\
In this section, we first discuss the classification based on the initial attractors, analyze the dynamics underlying these three types and discuss how they may be classified at later times. Animated versions of the orbits in PC space for all types (Fig. \ref{fig:A_Orbits}, \ref{fig:B_Orbits}, \ref{fig:CAttractor}), as well as the plots for Fig. \ref{fig:CFixedPoint} are available in our \href{https://github.com/DaveyPlugers/Evolution_Of_Robust_Cell_Differentiation_Under_Epigenetic_Feedback}{public GitHub repository}.
Finally, we discuss the occurrence of these processes and their reliance on the speed of epigenetic feedback $\nu$. Due to the high dimensionality ($M=40$) of gene expression level dynamics, we frequently use Principal Component Analysis (PCA) to visualize the orbits and differentiation process. (Note that 60-90\% of the variance of $x_i$ are captured by three PCs in the examples below, whereas any discussion on fixed-points are those in the M-dimensional (M=40) dynamical system.)

%% file: 3.TTM/Mechanisms_Classification2.tex
%The classification of these networks is based on the initial dynamics and their accompanying bifurcation mechanism that leads to differentiation. These mechanisms were initially identified by distinguishing between networks where gene expression levels $x_i$ are fixed (type B) and those where they are actively changing due to chaotic or periodic attractors and transient orbits (types A and C). This secondary group was then subdivided into networks with chaotic gene expression dynamics (type A) and networks with diverging orbits that are attracted to fixed points or non-chaotic limit cycles (type C). For type C, these two seemingly different mechanisms were combined due to both exhibiting transient, diverging orbits and an abundance of basins of attraction to fixed points.\\
%\\
The three types are classified by the attractors of the initial gene expression dynamics at $\theta_i = 0$. These initial attractors are quantitatively identified by calculating the Lyapunov exponent $\lambda$ for gene expression dynamics with $\theta_i$ fixed at 0. The three types are classified as:
\begin{itemize}
    \item Type A: Chaotic attractor ($\lambda>0$)
    \item Type B: Weakly stable fixed points ($x_i\approx0$ and $-1 < \lambda < 0$ )
    \item Type C: Stable attractors in the form of limit cycles ($\lambda=0$) or highly stable fixed points ($\lambda\approx-1$)
\end{itemize}
Depending on the initial attractor, distinct bifurcations occur, which are responsible for the multiple cell fates of the multicellular system. Therefore, the initial dynamics function as an appropriate proxy in determining the differentiation process allowing us to distinguish between the three types clearly. Before explaining each type in depth in later sections, we briefly summarize the behavior of each type below.\\
\\
Type A networks undergo a reduction of dimensionality for their attractors to go from chaotic attractors to fixed points, and they are classified by their positive Lyapunov exponent ($\lambda>0$). The change in $\theta_i$ instigates this reduction from chaos to limit cycles to fixed points. The state of $x_i$ determines the change in $\theta_i$, leading to the succeeding limit cycles being separated into varying phase spaces that define multiple subcycles. These are then further reduced to fixed points that represent the final cell state. In this hierarchical subdividing, orbits are successively separated, and the dimensionality of the network gradually decays.\\
\\
For type B, the initial fixed point includes some $x_i$'s with $x_i\approx0$, far from the fixed points at $\pm1$ thus the fixed point is weakly stable. The differentiation occurs through a slow migration of the fixed point, either $x_i=+1$ or $x_i=-1$, induced by noise and the change in $\theta_i$ to $+1$ or $-1$ respectively. Then, bistable states in $\theta$ are generated.\\
\\
For type C, the initial attractor is either a limit cycle ($\lambda=0$) or a stable fixed point for which all $x_i\approx\pm1$. Through the change in $\theta$, multiple fixed points are generated by saddle-nodes or SNIC bifurcations.\\
\\
Since the attractors of gene expression dynamics are classified to chaos, limit cycle (or quasiperiodic state), and weakly or fully stable fixed points, the classification is widely applicable and generally valid across models with different parameter values. Still, while networks can be distinguished through their initial attractor, there were some networks whose behavior changed as the epigenetic factors $\theta_i$ evolved. This causes networks to differentiate in multiple stages with different types (see Fig. \ref{fig:Combi_Oribts} in the Appendix). We do not use such networks in our analysis, and we do not regard this as a fourth type since it is possible to recover the three types by considering these as distinct sequential differentiation processes.
%Since the classification is based on initial attractors, we can confidently state that types A and C account for all the well-established attractors in dynamical systems theory (chaos, limit cycles, and fixed points). However, type B introduces a novel mechanism of directed diffusion, suggesting the possibility of other undiscovered mechanisms. While no additional mechanisms have been identified, nor could we imagine any, we focus our discussion on the three observed types without ruling out that others may arise. We summarize the mechanisms in Table \ref{Tabel_ref} and discuss each in detail in their respective sections. ...... Discuss, I can either discuss more here on the continued time course and remove section D. Or write in section D on this, although it may not be very long?

\renewcommand{\arraystretch}{1.4}
\begin{table*}[ht]
\centering
\begin{tabularx}{\textwidth}{ 
  | >{\centering\arraybackslash}p{1.2cm}
  | >{\centering\arraybackslash}X 
  | >{\centering\arraybackslash}X 
  | >{\centering\arraybackslash}X 
  | >{\centering\arraybackslash}X | }

\hline
\textbf{Type} & \textbf{Criteria by initial attractors at $\theta_i=0$} & \textbf{Time Course} & \textbf{Differentiation Process in terms of Dynamical Systems} & \textbf{Noise and I.C. Dependence} \\
\hline
A & Chaotic orbits ($\lambda_{\text{Lyap}} > 0$) & Reduction from chaos to oscillation to fixed point & Hierarchical branching of chaos to subcycles and to fixed points & Robust to noise and initial perturbations \\
B & Weakly stable fixed points ($-1 < \lambda_{\text{Lyap}} < 0$) and frustrated fixed points ($x_i^* \sim 0.$) & Slow change in fixed points towards the direction with higher stability and eventual stable final state & Bifurcation of fixed points by directed motion with epigenetic fixation of noise-induced perturbations & Positive role of noise for differentiation; robust to initial perturbations. \\
C & Limit cycle ($\lambda_{\text{Lyap}}=0$) or stable fixed points ($\lambda_{\text{Lyap}}=-1$)& Sudden switch to multiple different fixed points & SNIC or Saddle-Node bifurcation & Not robust to perturbations; high noise leads to loss of differentiation. \\
\hline
\end{tabularx}
\caption{Summary of the three types and their differentiation properties. }
\label{Tabel_ref}
\end{table*}

%% file: 3.TTM/Mechanisms_A.tex
The first type appears in both low- and high-noise simulations. It is characterized by long-lasting oscillatory dynamics that remain even in the absence of noise. Initially, cells enter a chaotic attractor, but as epigenetic modification levels ($\theta_i$) change, their orbits transition into periodic attractors. Small noise-induced variations are amplified by chaotic dynamics, causing cell orbits to diverge. Once gene expression level dynamics settle into cycles, continued epigenetic fixation leads them to distinct fixed points, depending on their expression history.\\
\\
\begin{figure}[!h]
    \centering
    \includegraphics[width=1\linewidth]{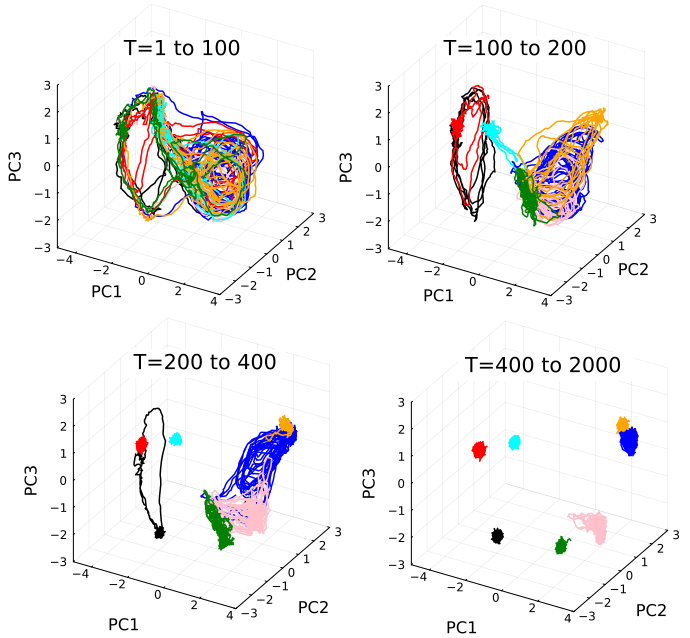}
    \caption{Orbits in PC space at various stages of differentiation for a given evolved network. The color indicates the different final cell types. PCs are obtained from $x_i$'s for the developmental time $T=1$ to 2500. Orbits of $x_i(t)$'s are plotted by three PCs for the displayed ranges of developmental time. These time ranges represent the different developmental stages. This specific example finishes type A differentiation much faster ($T\approx400$) when compared to other type A processes ($T\approx1000$) but is chosen for its clearer visualization. The initial shared periodic attractors diverge and they eventually reach fixed point states. We verified these were fixed points in the 40-dimensional phase space of gene expression levels as confirmed by their time-series.}
    \label{fig:A_Orbits}
\end{figure}\\
Fig. \ref{fig:A_Orbits} demonstrates example orbits from a type A network. Up to $T=100$, all cell types share a chaotic attractor. At $T=100$, orbits begin to diverge, and cells settle into two distinct cycles. Here, noise plays a supportive role; small differences introduced by noise are amplified by the instability in chaotic dynamics. This amplification shifts the cycles, with epigenetic modifications stabilizing the displacement. By $T=200$, orbits are fully separated into two distinct regions. Over time, orbits undergo hierarchical separation due to epigenetic modifications. The red and black cell types remain in the leftmost orbit, while the orbits on the right have begun to separate. At this time, some cells exhibit fixed gene expression levels, while others continue oscillating. Over time, continued changes in epigenetic modification $\theta_i$ lead all the cycles toward fixed points. By $T=400$, epigenetic modifications stabilize cellular states into their final differentiated forms, preventing noise from disrupting the fixed points.\\
\\
\begin{figure}[!h]
    \centering
    \includegraphics[width=1\linewidth]{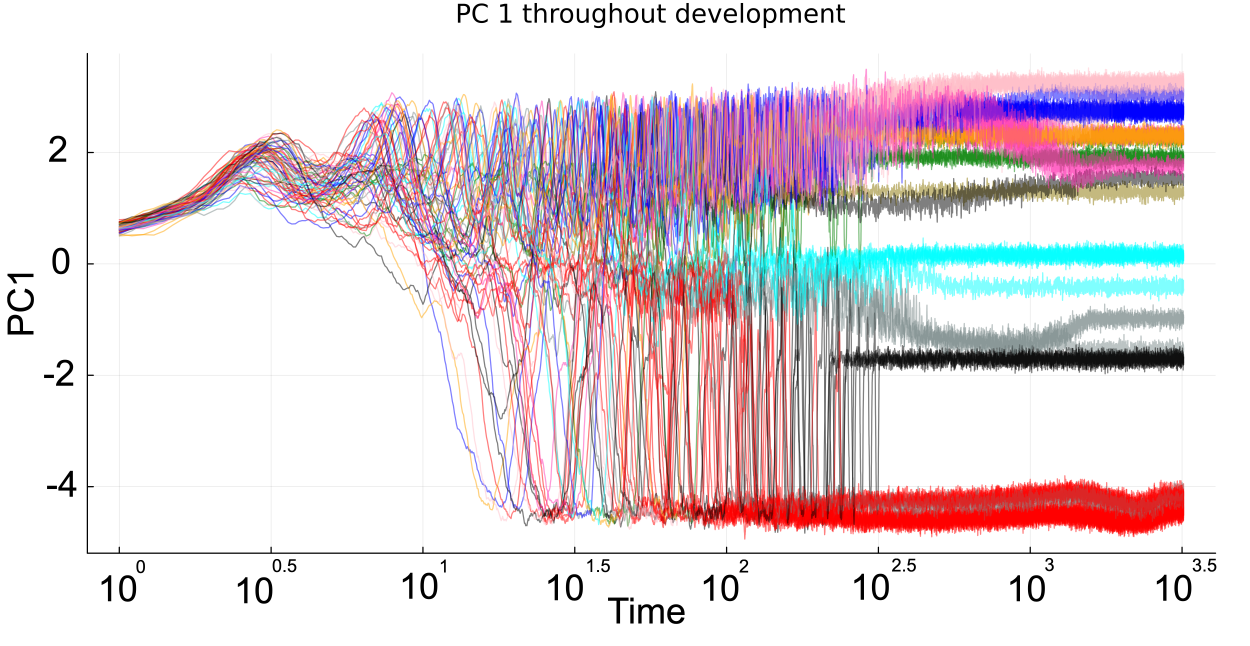}
    \caption{The time course of the first principal component throughout differentiation for a type A network. Initial dispersion ($T\approx 10^{1.3}$) is followed by cells sharing and jumping between subcycles ($t=10^{1.5}-10^{2.5}$) to eventual fixation once these cycles have died out ($T=10^{3}$). The simulation continues until $T\approx 10^{4.3}=20000$ but no longer shows any changes. The color indicates the different final cell types.}
    \label{fig:A_PCInTime}
\end{figure}\\
\\
\newpage
This entire progression, from chaotic dynamics to shared cycles, sub-cycles, and eventual fixation, is discernible in the time series of principal components in Fig. \ref{fig:A_PCInTime}. Notably, since only the first four genes determine cell type, different fixed points can occasionally correspond to the same cell type, as seen for cyan and gray.\\
\\
To see how the gene expression dynamics change with the slow change in epigenetic modifications, we freeze the epigenetic factors ($\theta_i$) at specific time points and analyze the resulting gene expression dynamics. This approach makes it possible to characterize the dynamics of gene expression levels $x_i$ for given $\theta_i$. Fig. \ref{fig:TypeA_FrozenTheta} illustrates the various stages of the differentiation process. At $T=50$, the dynamics exhibit a chaotic attractor, causing gene expression levels to disperse. As we take $\theta_i$ at later times, the orbits split apart from the chaotic attractor for $T=250$ and undergo further differentiation at $T=650$ (see black + green on top and pink + blue + maroon on the left). Finally, the system stabilizes at later times ($T=950$), although some gene expression levels continue to fluctuate. A more detailed quantitative analysis of this process is given in section \ref{sec:Attractor}.\\
\\
\begin{figure}[!h]
    \centering
    \includegraphics[width=1\linewidth]{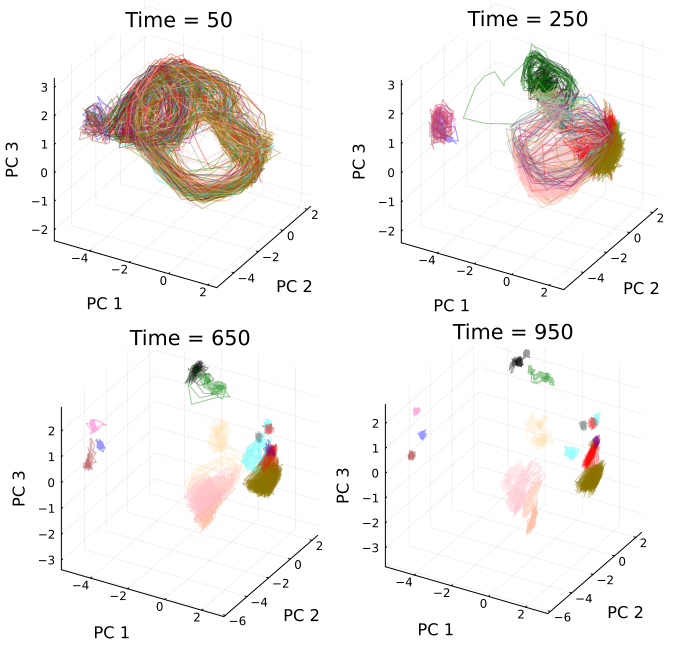}
    \caption{Attractors for the genes $x_i$ are displayed in PC space for fixed $\theta_i$ at the given times. The color indicates the different final cell types as defined by the target genes. By setting $\theta_i$ as a constant, we can analyze the full orbit of $x_i$ and distinguish these dynamics from the change in $\theta_i$ during development. Trajectories are generated by assigning each cell the gene expression levels $x_i$ and epigenetic factors $\theta_i$ exhibited by a cell at the given times as it goes through a differentiation simulation. (This plot shows a different example from Fig. \ref{fig:A_Orbits} and \ref{fig:A_PCInTime})}
    \label{fig:TypeA_FrozenTheta}
\end{figure}\\
\\
Note that at the last stage of differentiation into fixed points, epigenetic fixation by noise as in type B or saddle-node bifurcation as in type C sometimes follows. These networks exhibiting a secondary differentiation processes have been excluded from the analysis for simplification.

%% file: 3.TTM/Mechanisms_B.tex
\begin{figure}[!h]
    \centering
\includegraphics[width=1\linewidth]{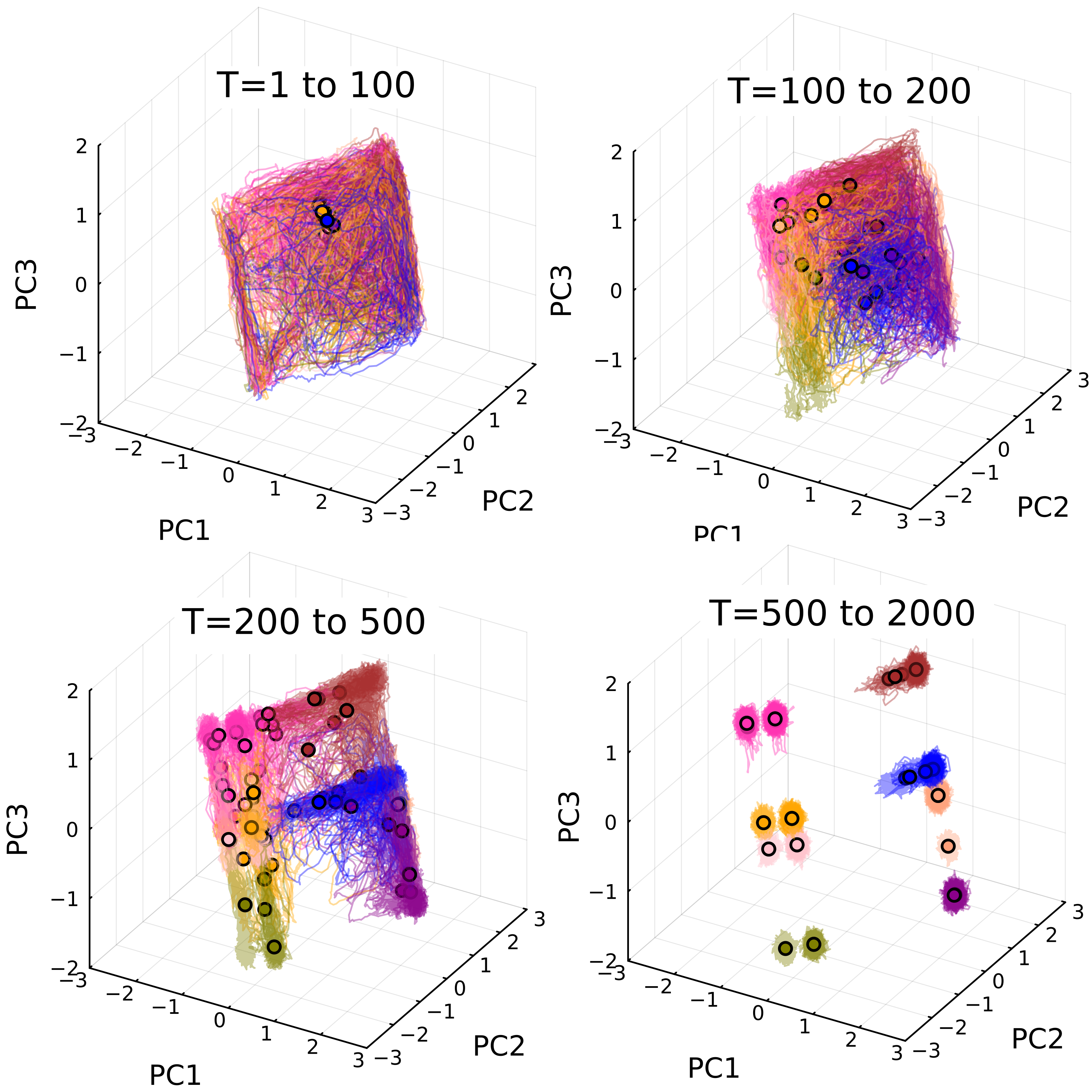}
    \caption{Orbits in PC space at various stages of differentiation. The color indicates the different final cell types. The time ranges represent developmental stages, and their exact values differ from network to network. The dots represent the fixed points at the start of the snapshot in which the cells will fall if the noise is turned off and the epigenetic modifications are kept constant.}
    \label{fig:B_Orbits}
\end{figure}

The second type mostly occurs in high-noise conditions and is infrequent in low-noise cases. Here, when noise and epigenetic modifications are removed, gene expression levels converge to fixed points and remain stable. With noise present, these fixed points migrate and diverge, leading to distinct final states. The epigenetic modifications $\theta_i$ determine the position of these fixed points. This migration is constrained to a low-dimensional space, as illustrated by the PC visualization in Fig. \ref{fig:B_Orbits}. \\
\\
\\
\\
Differentiation follows a highly structured pattern along parallel channels in the PC space of gene expressions, contrasting with type A, which exhibits chaotic oscillations and varies significantly across generations. For type B, the initial transient dynamics are brief, and cells quickly settle into fixed points. With noise and epigenetic dynamics turned off, cells are attracted toward a weakly-stable fixed point. Near this fixed point, small amounts of noise are sufficient to cause gene expression levels $x_i$ to diffuse. With epigenetic modifications, the fixed points shift gradually within a constrained low-dimensional space. \\
\\
An example of this differentiation process is shown in Fig. \ref{fig:B_Orbits}. For $T=1-100$, all fixed points are initially clustered together while orbits, influenced by noise, explore a large region of space. Around $T=100$, the stable region in which the gene expressions of cells exist begins to shift, and cells with similar final fates cluster together. This shrinking of the stable regions is due to the change in $\theta_i$, driving the fixed points of cells to migrate.
By $T=200$, the stable regions have shrunk considerably, forming a frame-like structure. These parallel lines are the channels through which the fixed points of each cell migrate to their final position. Finally, at $T=500$, the cells have mostly settled down into their final states. \\
\\
\begin{figure}[!h]
    \centering
    \includegraphics[width=1\linewidth]{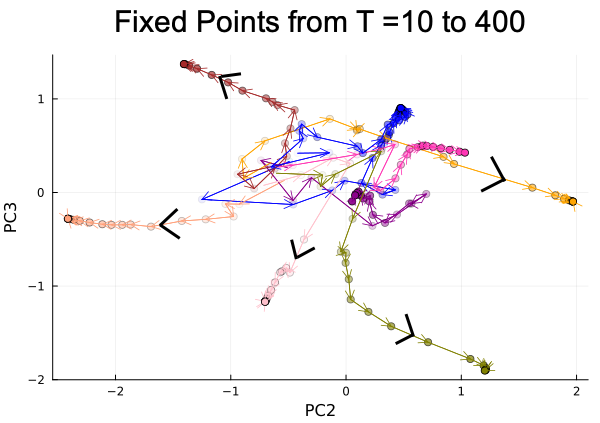}
    \caption{Orbits of the fixed point for various final cell types. Each cell has a distinct fixed point which directionally slowly moves due to the change in epigenetic modification. The orbits initially start close to the center before diverging out into parallel channels.}
    \label{fig:B_fixed points}
\end{figure}\\
This fixed-point migration was confirmed by disabling noise at various snapshots and tracking fixed points as the epigenetic modifications changed. The fixed points, corresponding to their current epigenetic modifications, move over time through the channels toward their final state, as shown in Fig. \ref{fig:B_fixed points}. These fixed points are also shown in Fig. \ref{fig:B_Orbits}, where one can observe how the noisy cell orbits are located near and dragged along with the fixed points. \\
Interestingly, rather than creating multiple fixed points as $\theta_i$ changes, only a single fixed point is present for each cell. The absence of other fixed points was verified numerically by taking snapshots in our simulation, removing all noise, freezing, and setting the epigenetic factors to those of a random cell. Independently of the current cell position or snapshot time, all cells converge to the same fixed point. (See Fig. \ref{fig:BSingleEpi} in the Appendix) \\
\begin{figure}
        \includegraphics[width=1\linewidth]{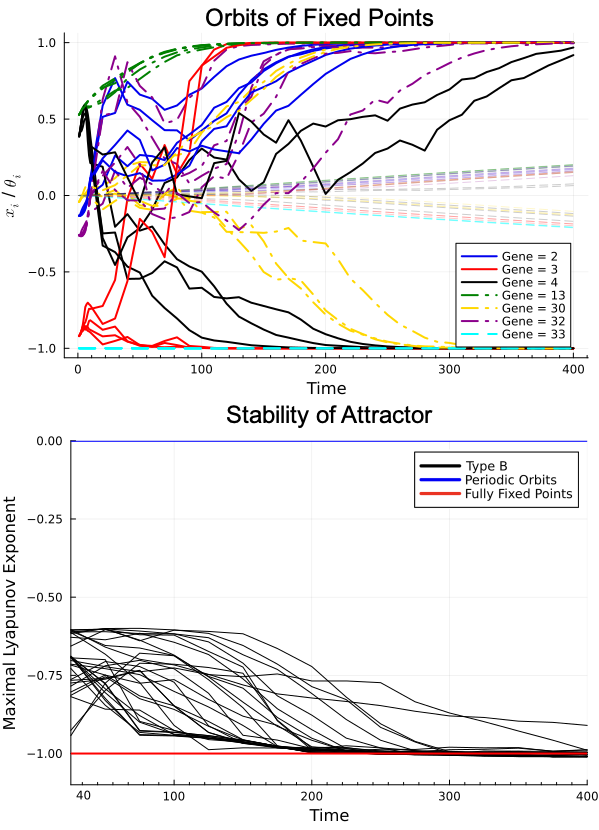}

    \caption{Upper: The time course of gene expression level $x_i$ corresponding to targets (full), for non-targets (dash-dot) and epigenetic factors $\theta_i$ (dashed + transparent) at successive time snapshots of the simulation with frozen $\theta$ and noiseless orbits. Lower: Values of maximal Lyapunov exponent throughout development indicating the presence of a stable state that is not a fixed point. As a reference, the blue and red lines correspond to the exponent for a periodic limit cycle and a fixed point at $x_i = \pm 1$ respectively.}
\label{fig:fixed points_FewGenes_B}
\end{figure}\\
\noindent
How do fixed points exhibit directional motion despite noise-induced diffusion? Note that, during the differentiation process, fixed points of $x_i$ are not close to $\pm 1$ for some genes i, as shown in the upper panel of Fig. \ref{fig:fixed points_FewGenes_B}. We refer to these as frustrated fixed points, as the value of some gene expressions is stuck in an intermediate state around $x_i^* \sim 0$, and requires the fixation through epigenetic modification to become fully stable. In fact, the maximal eigenvalue of the Jacobi matrix is larger than -1 which would be the expected value\footnote{$\partial_j (dx_i/dt) \approx 0-\delta_{ij}$ when $|Y| > 0.15$ for  $tanh(\beta*Y) \Rightarrow Jacobian \approx -I_M$} for the stabilized fixed points $x_i \in \{-1,1\}$, as seen in the lower panel of Fig. \ref{fig:fixed points_FewGenes_B}. This suggests that stability is weaker compared to fixed points in on/off states. As these $x_i$ values of the fixed point move towards $\pm 1$, stability increases, causing the influence of noise to diminish. Thus, the directional motion of fixed points with $x_i = \pm 1$ follows. The fixed points move along the eigenvectors corresponding to eigenvalues larger than -1, influenced by noise and epigenetic modifications, leading to the frame-like structure seen in Fig. \ref{fig:B_Orbits}. It also demonstrates that the timescale over which these genes change can vary widely depending on the genes. For example, the expression of genes 3 and 13 approaches $\pm 1$, while genes 4, 30, and 32 remain at intermediate values between $\pm 1$, demonstrating much slower dynamics.

%% file: 3.TTM/Mechanisms_C.tex
\begin{figure}[!h]
    \centering
    \includegraphics[width=1\linewidth]{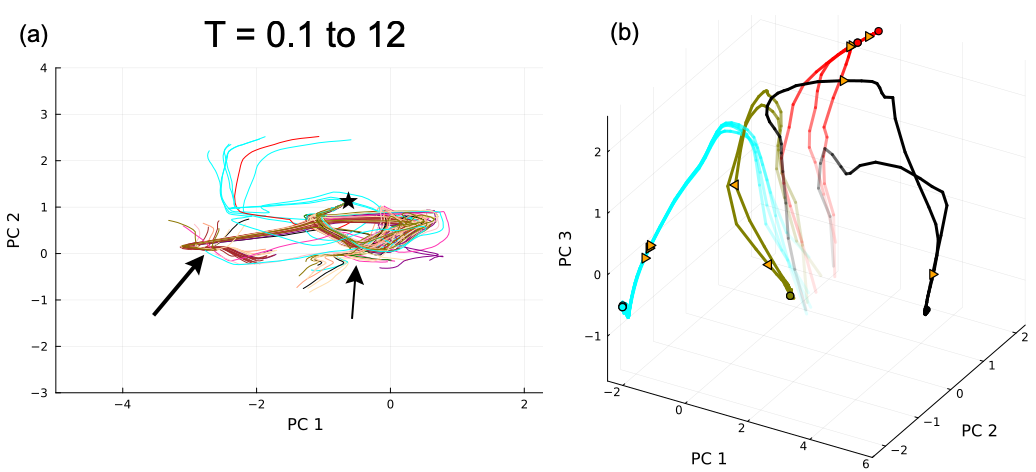}
    \caption{Orbits of quenching fixed point type C processes with color used to indicate the final cell types. (a) Early gene expression dynamics start at the star and pass by the saddle points (-0.5,0) and (-3,0) along the unstable manifolds. (b) The orbits of gene expression levels with perturbed initial conditions and where noise is removed. Orbits move towards multiple fixed points with the orange arrow indicating the direction and position of the cells at $T=10$, circles represent the final fixed point. Animated versions of these plots are in our \href{https://github.com/DaveyPlugers/Evolution_Of_Robust_Cell_Differentiation_Under_Epigenetic_Feedback}{public GitHub repository}.}
    \label{fig:CFixedPoint}
\end{figure}
\begin{figure}[!h]
    \centering
    \includegraphics[width=1\linewidth]{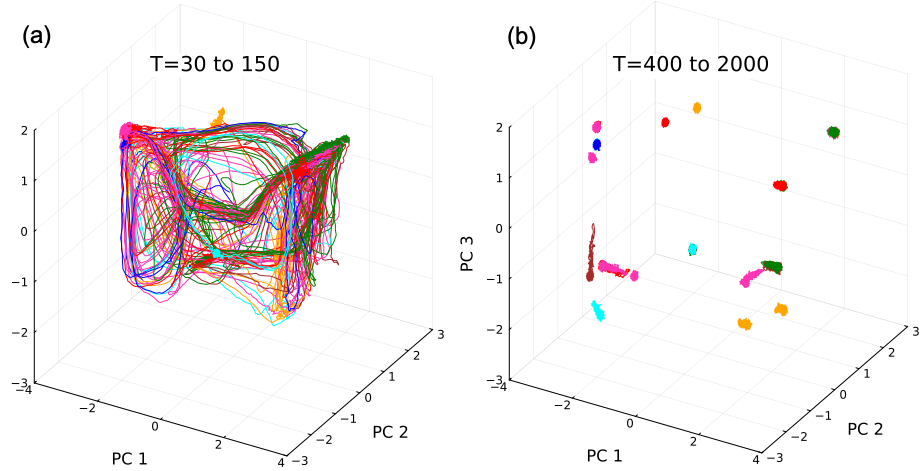}
    \caption{Orbits of quenching attractor type C mechanisms with color used to indicate the final cell types. Left: Gene expression dynamics before quenching. The initial attractor halts around $T\approx150$ and transitions into stable fixed points and transient orbits for $T=150-400$. Right: All cells have stabilized in fixed points with some slowly shifting positions (maroon, pink).}
    \label{fig:CAttractor}
\end{figure}

The type C quenching process is most commonly observed for the lower noise case. The cells quickly approaches stable fixed points after passing a saddle point or a limit cycle, where they undergo a sudden cell fate decision process without successive reduction in the oscillation (type A) or slow migration of fixed points (type B). After initial transient oscillations, several stable fixed points, separated by saddle points, appear within a short time interval, and the cellular states are attracted to each of them and fixed. 
There are two ways to achieve different fixed points.\\
\\
\\
First, when the cellular state crosses a saddle point by noise, slight differences in gene expressions cause the cellular state to be on opposite sides of the unstable manifold of the saddle point, from which cells are led to distinct final fates. In the second version, periodic attractors collide with the saddle points, and thus, once the periodic attractor disappears, orbits are attracted to novel fixed points.\\
\\
Fig.\hyperref[fig:CFixedPoint]{\ref{fig:CFixedPoint}.(a)} illustrates a type C differentiation process with quenching fixed points. The initial orbits, starting from the star, lead the cells through 2 saddle points indicated by the arrows. Passing the first saddle point at (-0.5,0), one set of orbits moves to the second saddle point at (-3,0), while other orbits go up (cyan, red) and down (green, pink, black, yellow). Then the orbits are separated by the second saddle point, either moving to the right (maroon) or dispersing orbits going up (green, pink, orange, black, maroon). The orbits, now sufficiently spread apart, continue their dynamics and quickly fall into a nearby fixed point that will become the final cell state. The presence of these fixed points and the attraction to each of them are demonstrated on the right Fig.\hyperref[fig:CFixedPoint]{\ref{fig:CFixedPoint}.(b)}. Here, the noise is set to 0, and the initial conditions are perturbed to induce different cell fates. Cells that initially undergo similar dynamics are quickly spread apart to distinct cell fates following the unstable manifold of saddle points.\\
\\
Fig. \ref{fig:CAttractor}
shows the dynamics of the quenching periodic attractor process. The initial periodic attractor, shown in Fig.\hyperref[fig:CAttractor]{\ref{fig:CAttractor}.(a)}, is maintained until the orbits suddenly quench to fixed points. This quenching allows cells to fall into fixed points or undergo transient dynamics before settling down, as seen in Fig.\hyperref[fig:CAttractor]{\ref{fig:CAttractor}.(b)}.\\

\begin{figure}[!h]
    \centering
    \includegraphics[width=1\linewidth]{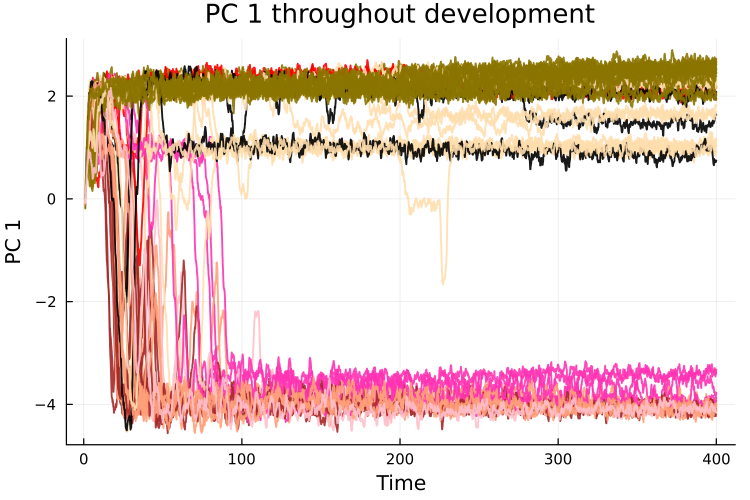}
    \caption{The time series of the first principal component throughout differentiation for a type C network. 64 cells from the same initial conditions are plotted and demonstrate rapid spreading and commitment to their final cell fate at $T=50$.}
    \label{fig:PCA_C}
\end{figure}

The time course in type C is distinguishable from types A and B. In type A, complex orbits are successively replaced by simpler, smaller-scale limit cycles, whereas in type C, the periodic orbit suddenly terminates in a short time span and is replaced by fixed points. These fixed points are sufficiently stable so that they do not diffuse by noise in contrast to type B.\\ 
\\
Fig. \ref{fig:PCA_C} demonstrates the time course of differentiation by using the first principal component for Fig. \ref{fig:CFixedPoint}.a. After the initial dynamics, most cell fates become fixed after T=80. Some cells still exhibit some oscillation after cell fate commitment, but these decay due to the change in epigenetic modifications that freeze out all gene expression levels at later times. Quenching from oscillatory states gives a similar PC time series where they exhibit global oscillation before quenching.\\
\\
Differentiation for type C is sometimes accompanied by the noise-induced epigenetic fixation in type B. For instance, starting with a type C which crosses saddle points and leads to branched orbits. Later in the differentiation process, they adopt weakly stable fixed points from which a secondary differentiation follows. Fig. \ref{fig:Combi_Oribts} in the Appendix shows the 2D PC dynamics of such a network.

%% file: 3.TTM/Mechanisms_Statisitcs.tex
\begin{figure}[!h]
    \centering
    \includegraphics[width=1\linewidth]{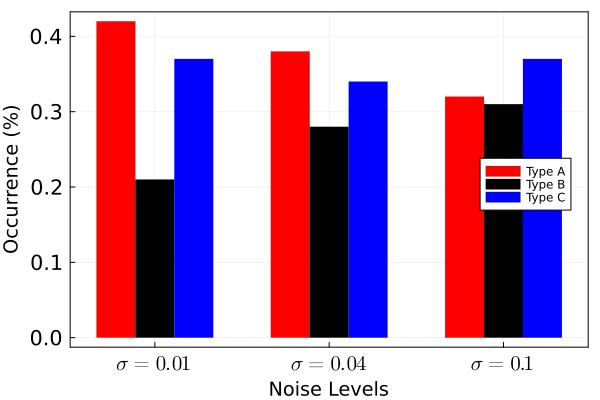}
    \caption{Occurrence of the three types of networks analyzed for 50 networks with M=40 evolved under low, medium, and high noise. Networks are identified through the Lyapunov exponent at $\theta_i=0$. Networks that have $\lambda_{Lyap}>0.001$ are classified as type A. Networks that have $\lambda_{Lyap}\in[-0.97,-0.15]$ are considered to be type B. Otherwise, it is a type C network with $\lambda_{Lyap}\in[-0.15,0.001]$ or $\lambda_{Lyap}<-0.97$. Type B networks favor high noise environments as it makes differentiation more feasible. }
    \label{fig:ABCStat}
\end{figure}

We examined the fractions of type A, B, and C by 50 samples randomly chosen from the evolution by using the Lyapunov classification scheme. We also explore how these fractions change with the noise levels $\sigma \in \{0.01,0.04,0.1\}$. Fig. \ref{fig:ABCStat} demonstrates that the fraction of type B networks decreases for lower noise levels, whereas types A and C become more common. This trend is caused by the fact that type B networks rely strongly on noise to achieve differentiation, in contrast to types A and C, where internal dynamics allow differentiation at lower noise levels.\\
\\
\\
\\

%% file: 3.TTM/Mechanisms_nu.tex
So far, we adopt the case with $\nu=6\cdot10^{-4}$, implying that the epigenetic modification is sufficiently slow. Now we examine the dependence of the results on this epigenetic timescale $\tau_\theta = 1/\nu$. Samples that were evolved for $6\cdot10^{-4}$  are examined whether their cell differentiation still works when $\nu$ is increased up to $10^{-1}$. First, for type A, almost all of the differentiated cell types are retained up to $\nu=10^{-2}$. Increasing to $\nu = 10^{-1}$ reduces the number of recovered cell types but still retains the differentiation process.\\
\\
For type B, the amount of recovered cell types is halved for $\nu=10^{-2}$, and any form of differentiation is lost as $\nu$ approaches $10^{-1}$. For type B, the increase in $\nu$ is equivalent to increasing the speed of cells in the differentiation channels, causing the cell fate valleys to split apart rapidly. Due to this rapid cut-off, cells will no longer have the time to diffuse in their channels to different final cell fates, leading to a reduction in the number of final cell types. \\
\\
The type C is more vulnerable to the increase in $\nu$. Some networks show the loss of some cell types at $\nu=10^{-2}$, while other networks already lose some for $\nu = 10^{-3}$. This is probably explained as follows: The basins of attraction for the fixed points increase due to the change in epigenetic modification. When $\nu$ is increased, some of them grow much faster and catch cells during transient dynamics, eliminating the attraction to some other types.

%% file: 4.Analysis/Analysis_0Intro.tex
In this section, we make a quantitative analysis to understand and distinguish the three types of differentiation processes. 

%% file: 4.Analysis/Analysis_1Attr.tex
\begin{figure}[!h]
    \centering
    \includegraphics[width=1\linewidth]{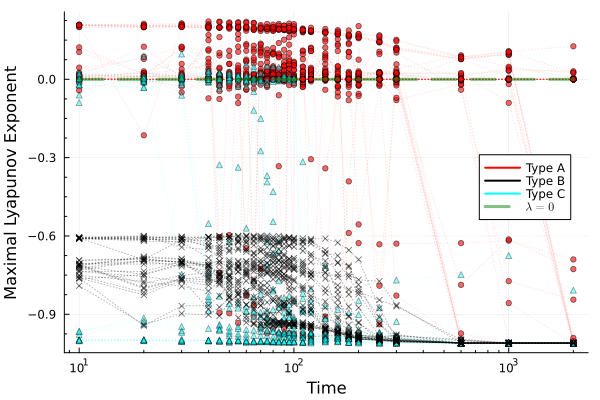}
    \caption{Maximal Lyapunov exponents plotted for 3, 3, and 4 type A, B, and C networks respectively. Each network has the exponents of 10 cells plotted to sample the behavior. Type A networks exhibit positive values indicating a high degree of chaos while type B is always negative and stable. Type C networks exhibit negative and 0 for their exponents due to their dynamics consisting of fixed points and periodic cycles.}
    \label{fig:A_Lyap}
\end{figure}

We examine the attractors of gene expression dynamics by fixing the epigenetic modification $\theta_i$ and study how their fraction changes with the temporal change in $\theta_i$ through the course of differentiation. First, we distinguish the fixed-point and the non-fixed-point attractors. The latter is either (quasi-)periodic or chaotic, which can be distinguished by computing the Lyapunov exponent and judging if it is positive or not.\\
\\
For the analysis, the gene expression dynamics are run without noise and by fixing $\theta_i$, after sufficient transient steps to allow for the cell to reach its attractor. We examine if the attractor is time-varying or a fixed point by computing the following quantity $\alpha$:
\begin{equation}
    \alpha = \int_0^{\tau} \sum_{i=1}^M \left \vert \frac{dx_i (t)}{dt}\right \vert dt
\end{equation}
The state is considered to be in a fixed point if $\alpha \approx 0$ (Numerically $\alpha/\tau < 10^{-4} $). 
Next, we compute if there are multiple fixed points and count their number. This is achieved by computing the Euclidean distance of fixed points, and if this distance is beyond 0.1, it is regarded as a distinct fixed point. To examine if the dynamics are chaotic or not, we compute the maximum Lyapunov exponent for the gene expression dynamics while fixing $\theta_i$ and removing noise. We adopt the standard algorithm by using the Jacobian matrix, and convergence of the Lyapunov exponent is verified with the moving average.\\
\\
\begin{figure*}
    \centering
    \includegraphics[width=0.9\linewidth]{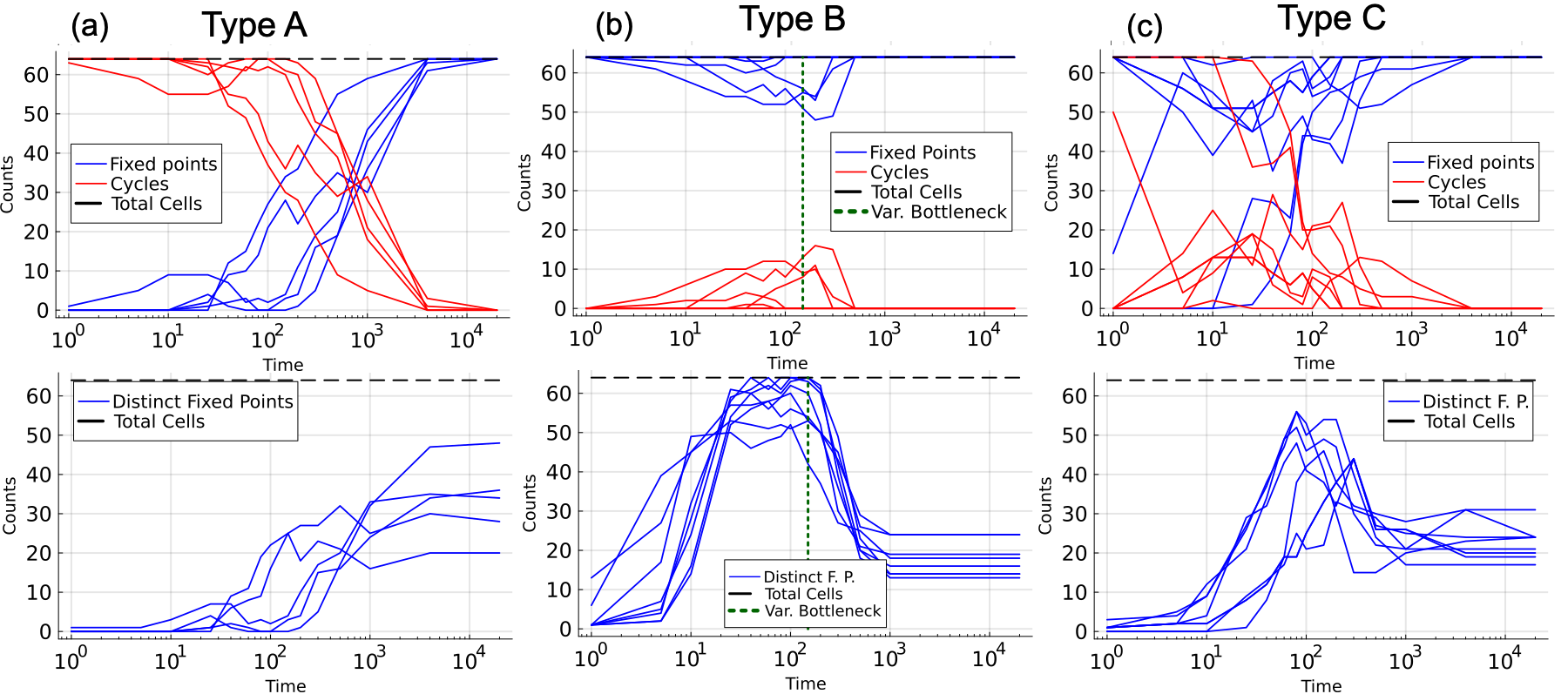}
    \caption{Attractor count (y-axis) of type A, B, and C networks determined by freezing out the epigenetic factors at various times (x-axis). The upper plots display the number of cells that fall into oscillatory attractors or a fixed point. The lower plots show the number of distinct fixed points, where cells that fall in the same fixed point are only counted once. The figures plot the results for 5, 8, and 7 networks for types A, B, and C, respectively. For type A, all cells are initially in cycles; as the simulation progresses, the change in epigenetic factors replaces cycles with fixed points starting around $T=10^2$. For type B, virtually no cycles exist, but the amount of fixed points varies drastically throughout the simulations. The initial fixed points have an early divergent period where they spread apart, leading to the peak at $T=10^2$. Once channelization occurs, these fixed points approach and merge, leading to the decline in distinct fixed points. We have marked the end of the variance bottleneck (see \ref{subsec:Var}) as this coincides with fixed points channelization toward the final cell state. For type C, cycles may be present dependent on the version, but in much lower quantities and dying out much earlier than type A. Cells fall into fixed points between $T=10^1$ and $T=10^2$ and merge around $T=4 \cdot 10^2$. These mergers are caused by fixed points of similar cell types that are still approaching a common final state.}
    \label{fig:Attractors}
\end{figure*}
We plot the time course of the fraction of each attractor type for A, B, and C networks in Fig. \ref{fig:Attractors}. Initially, all the cells are in a dynamically varying, mostly chaotic, state for type A (see also Fig. \ref{fig:A_Lyap}). Starting at $T=10^2$, some cells start to fall into fixed points, and cycles disappear. In this stage, in the original simulation with noise, cellular states may not always remain at fixed points and may jump back into a cycle by noise. As time continues, all cycles are eventually eliminated and gradually replaced by fixed points. The number of distinct final fixed points over the cells increases monotonically with some sample dependence, as seen in the lower plot of Fig.\hyperref[fig:Attractors]{\ref{fig:Attractors}.(a)}. The large number of distinct fixed points is likely due to chaos and long-lasting cycles, allowing for more opportunities for cells to diverge. \\
\\
The fraction of fixed points in Fig.\hyperref[fig:Attractors]{\ref{fig:Attractors}.(b)} is almost unity from $T=1$, and there appear almost no oscillatory states. Rather, the attractor for given $\theta$ is only a single fixed point that moves slowly. As shown in the lower column, this fixed point is initially shared by all cells until the number of distinct fixed points suddenly increases at around $T=10^{1}$. The fixed points spread through the system so that the number of distinct fixed points reaches its maximum value. Later, due to the continued migration of fixed points, the number of distinct fixed points decreases as they approach the same final state and merge. Note that for some rare examples, a few cells experience cycles.\\  
\\
In Fig.\hyperref[fig:Attractors]{\ref{fig:Attractors}.(c)}, fixed points appear from early times, while some exhibit periodic states. The number of distinct fixed points shows a peak at $T=10^2$ and then slightly decreases. Some type C network cells fall into various fixed points, which then merge due to the epigenetic fixation, leading to the peak and decrease in the number of distinct fixed points similar to type B. Other type C networks initially show cycles before quenching and transitioning into fixed points. This is similar to type A, but in this case, the cycles are replaced by fixed points in a short time-span. This also leads to the observed peak for the number of distinct fixed points as all cells suddenly fall into fixed points and slow migration with mergers occurs slowly. This is in contrast to the monotonic increase of the number of distinct fixed points for type A.\\
\\
The properties of the attractors (chaotic, periodic, fixed points) are examined in Fig. \ref{fig:A_Lyap} by computing the maximal Lyapunov exponent throughout development for multiple cells in different networks. For type A, chaotic dynamics remain over the initial stage where the Lyapunov exponent remains positive. For some cells, the maximal Lyapunov exponent sometimes reaches zero, implying the (quasi)periodic behavior. Later, some cells sporadically take negative Lyapunov exponents. These negative exponents suggest the emergence of temporary or accidental fixed points, and finally, all cells take negative values when reaching their final fixed points.\\
\\
For type B, from the early stages, cells take negative Lyapunov exponents, indicating that the orbits reach a stable fixed point attractor. As already discussed, the computed exponent is initially much larger than -1, whose value is expected for $x_i = \pm 1$, indicating fully stable fixed points. The intermediate value of the exponent (around $-0.6 \sim -0.9$) slowly decreases toward -1. This demonstrates the presence of weakly stable fixed points and later moves to fully stable fixed points. \\
\\
Type C networks do not clearly show the existence of positive Lyapunov exponents. They instead have null exponents implying the existence of periodic attractors, followed by eigenvalues $\lambda \sim -1$, implying quenching to stable fixed points with $x_i  = \pm 1$. Intermediate Lyapunov exponents, similar to type B, are observed during transitions between periodic attractors and fixed points, but unlike B, these do not persist and immediately transition to $\lambda \sim -1$ instead.\\
\begin{figure}[!h]
    \centering
    \includegraphics[width=0.85\linewidth]{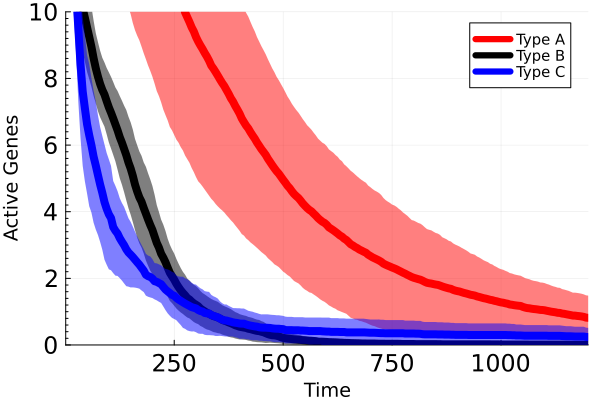}
    \caption{The average number of genes whose expression levels temporally change for 10, 9, and 4 type A, B, and C networks respectively. First, the temporal average of the actively changing gene expressions over all 64 cells is calculated for a single network. This is repeated for multiple networks that were evolved under the same noise strength. This average is plotted by thick lines, while the ribbon indicates the standard deviation between networks of the same type. Type C processes start with a very sharp drop off, with only a single non-fixated gene left at $T=300$, indicating rapid differentiation. Type A, however, still has roughly a quarter of its genes non-fixated at $T=300$ since it is still going through differentiation.}
    \label{fig:ActiveGenes}
\end{figure}\\
\\
\newpage
Another way to distinguish the three differentiation processes is to analyze the number of genes whose expression levels are not fully fixed at $x_i \sim \pm 1$. We define the threshold $|x_i| > 0.9$, and such genes $x_i$ are regarded to not be fully fixed. The number of such genes is plotted as a function of times for various networks in Fig. \ref{fig:ActiveGenes}. This shows the late fixation of type A networks compared to types B and C. 

%% file: 4.Analysis/Analysis_2Var.tex
To see how cell-cell variation changes through the cell differentiation process, we also computed the variance in the gene expressions across cells. We first compute global intercellular variance $\Gamma$ as the variance between all cells of all types, and then we compute the intra-cell type variance $\Gamma_{Intra}$ that is only across cells sharing the same final cell type. The former gives a degree of fluctuations across cells, leading to differentiation. While the latter gives the fluctuation within the same cell type. Low variance states indicate similarity in cells as there are fewer differences in the gene expression levels. To be specific, the above two variances $\Gamma$ and $\Gamma_{Intra}$ are defined as follows: By denoting the i-th gene expression of cell $j$ as $x_{i,j}$, the average expression of gene $i$ as $\overline{X_i}$ is defined by $\overline{X_i}(t) = \sum_{j=1}^N \frac{x_{i,j}(t)}{N}$. Then, the variance measures are computed through the standard variance as given by $\Gamma (t) = \sum_{i=1}^M \frac{\sum_{j=1}^{N} \left(x_{i,j}(t) - \overline{X_i}(t)\right)^2}{N-1}$.\\
\\
\begin{figure*}
    \centering
    \includegraphics[width=1\linewidth]{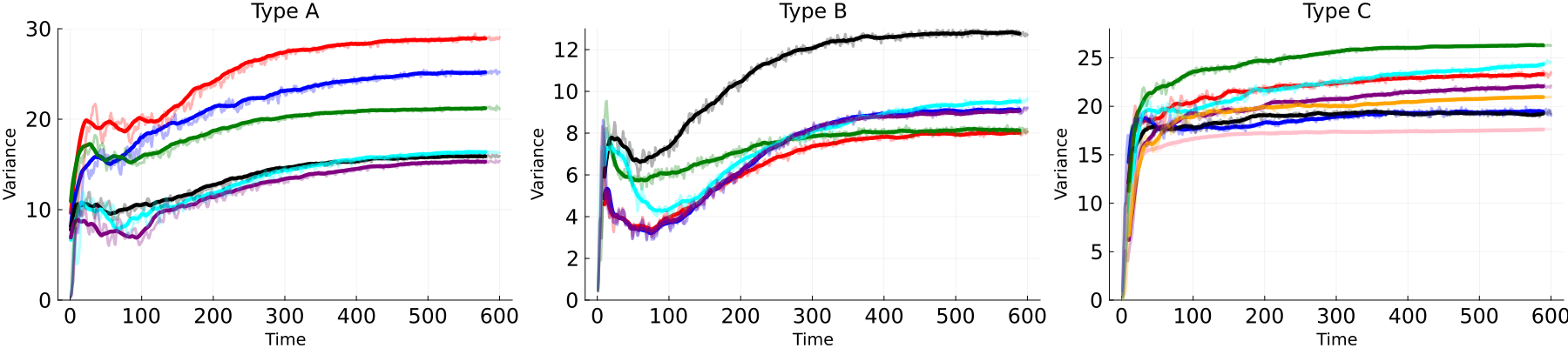}
    \caption{Interphenotypic variance for various differentiation processes. The variance $\Gamma(t)$ is computed over all $N=64$ cells. Thick lines give a time-averaged smoothed approximation of the exact values of $\Gamma$, which are plotted opaquely. Curves with different colors indicate results from different networks. All processes exhibit an initial peak but differ in their behavior afterward. Variance for type A gradually increases as orbits shift throughout the longer differentiation process. For type B, the initial peak gets reduced to a bottleneck as cells fall into their channels and increases again once these cells diverge apart toward their final states. Type C stays mostly the same, and most of the variance in these systems is obtained during the initial peak.}
    \label{fig:TimeCourseVariances}
\end{figure*}
The time course for $\Gamma(t)$ is shown in Fig. \ref{fig:TimeCourseVariances}. For type A networks, the variance has a fast initial increase due to chaos and then a later gradual increase throughout the differentiation. Type B networks exhibit a salient bottleneck in $\Gamma(t)$. Here, the variance initially increases (by noise) but is then followed by a strong drop (usually around 20-50\%) before increasing again throughout differentiation. This behavior for type B will be explained as follows: Initially, expression levels of all cells are close to each other, but by noise, the orbits are diversified. Shortly after this burst of the variance, it decreases as cells start to fall into their channels, which leads to the final cell stages. This channelization reduces the variance in gene expressions by cells, leading to the local minima around $T=100$. Finally, the epigenetic values start to change more, the steady state of the cells migrate away from each other and through this, the variance increases again.\\
\\
\begin{figure*}
\centering

    \includegraphics[width=1\linewidth]{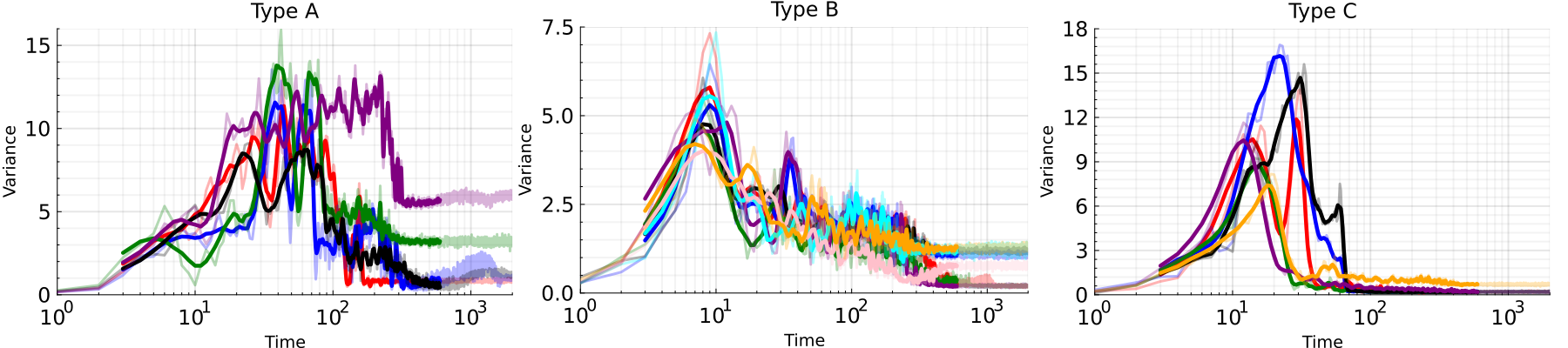}

    \caption{Intra-cell type variance for a single type A, B, and C network. One typical network per type is used, and the color indicates different final cell fates. All networks and phenotypes go through an initial higher variance state during differentiation. At later stages, this variance is reduced again as cells that will eventually share their phenotype approach each other.}
    \label{Fig:Intrapheno}
\end{figure*}
For type C, the variance shows a huge initial growth, as seen in type A. Shortly after this growth, they approach steady values for their variance. For some networks, secondary differentiation or a slow drifting of the final cell types leads to a small monotonic increase in the variance. The main difference from type A is that the increase in variance is rather small and is quenched at earlier times.\\
\\
So far, we have discussed the global variance over cells covering different cell types. Next, we study the intra-cell type variance between cells with the same final type, where the equations for $\overline{X_i}$ and $\Gamma(t)$ are adjusted with a filtered summation to only include cells that share their final phenotype. (This is done by summing over the subset of integers $C_j = \{k \vert p(j) = p(k)\}$ where $p(j)$ represents the cell type fate of cell $j$, the division of $N-1$ is also replaced with $\vert C_j \vert-1$.) This variance measure indicates the variance of gene expressions for given cell types and gives an estimate of when cells have committed to their final state.\\
\\
The time course of the intra-cell type variance is plotted in Fig. \ref{Fig:Intrapheno}. For all types A-C, there is an initial peak in the variance which decays back down as cells approach their final states. The differentiation processes have differences in their timing. For type A, the peak in variance tends to last for a long time before decaying ($T\approx 100$). For type B, there is a peak at earlier times ($T\approx 10$) where cells fall into differentiation channels. After this peak, the variance gradually decreases during the annealing process ($T= 10-400$). In contrast, type C has a very sharp peak and early drop ($T\approx 15$) as cells quickly approach their final cell state. In the discussion, we will compare this data with recently reported experimental results. 

%% file: 4.Analysis/Analysis_3Robust.tex
The robustness is a measure of the reproducibility of cellular states, given a constraint or disruption. Here we consider this robustness of the number distribution of cells of each type $i$ ($i=1,...,2^4$) with respect to perturbations of the initial conditions for the obtained GRN. \\
\\
First, we examine the number distribution of each cell type after the final cell types are reached at $T=20000$. Then, we repeat the simulations from perturbed initial conditions 10 times and examine the variation of the number distribution. This variation in cell type number is small for types A and B but has a large half-width for type C. Fig. \ref{fig:Boxplot} in the appendix shows an example of the distribution of the final cell types.\\
\\
To make a proper comparison, we quantify the robustness of the cell-type distribution. For this, we adopt the Kullback–Leibler Divergence (KLD), which measures the loss of information between two distributions of final cell type states under the application of perturbations. We define the vector $(\overrightarrow{X_\tau}) = (n_1,n_2,n_i,...,n_l)$ with $i$ representing cell type ($l=2^4$) and $n_i$ the number of cells with the i-th cell type generated from the initial 64 perturbed cells.
The average $\overline{\overrightarrow{X}}$ is computed according to $\overline{\overrightarrow{X}} = \sum_{\tau=1}^T \frac{\overrightarrow{X_\tau}}{T}$ where index $\tau$ represents different samples of $\overrightarrow{X_\tau}$ generated from 10 sets of perturbations with all 64 cells affected by the same perturbation. \\
\\
Then, KLD is defined by the equation: $KLD = \frac{1}{T} \sum_{\tau=1}^T \sum_{k=1}^{2^4} X_\tau[k] log\left( \frac{X_\tau[k]}{\overline{X}[k]} \right)$, where the summation over $\tau$ is taken across different samples with different initial conditions\footnote{It has also been verified that all the networks used in these runs consistently reproduce similar final cell states and have a low KLD value when not perturbed. This had to be done to not falsely ascribe the high KLD values to the perturbations when the underlying unperturbed network naturally causes this.}, and the summation over k represents the distinct final cell types (recall that there are 4 target genes and thus $2^4$ possible final states). Here, it is noted that the selection in evolution is based only on the number of existing cell types, while this robustness measure also takes the number distribution of each cell type into account. This robustness refers not only to the existence of the same cell types but also the number distribution of each type.\\
\begin{figure}[!h]
    \centering
    \includegraphics[width=1\linewidth]{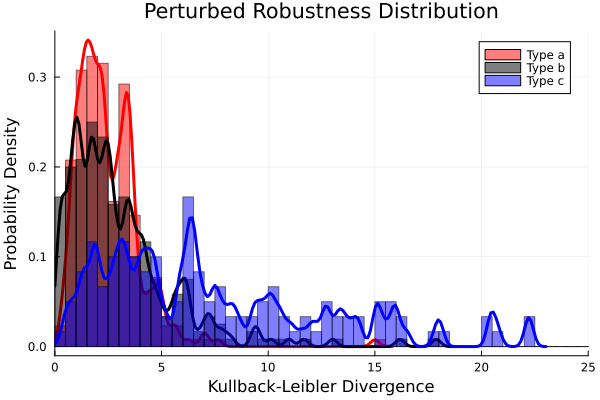}
    \caption{The values of KLD for 13, 15, and 9 A, B, and C networks respectively. They are obtained for the high noise case ($\sigma = 0.1$). Calculated through repeated simulations and perturbing the initial conditions for 10 runs per network by a radius of 5 ($\pm 0.79$ per $x_i$). The distribution for type A and type B networks mostly overlaps and consists of relatively low KL values, which makes them robust. Type C is much more spread out towards higher values, indicating their dependency on the initial conditions for differentiation.}
    \label{fig:Perturb_Rob}
\end{figure}\\
The KLD values of the multiple perturbed networks are calculated and binned together in Fig. \ref{fig:Perturb_Rob}. In type C, the KLD distribution is extended to large values, indicating that the number distribution of final cell types crucially depends on each sample. In contrast, for types A and B, the KLD distribution has a peak at a lower value and does not extend to higher values. Therefore, type A and B networks give rise to cell-type distributions robust against perturbations. In type A, attraction to distinct chaotic attractors over initial conditions is expected to lead to such robustness. In type B, generated fixed points are restricted into a common low dimensional space, independent of initial conditions. In contrast for type C, multiple fixed points are generated and which of them are normally reached through dynamics depends on initial conditions leading to the lack of robustness.

%% file: 5.Disc/0MainDisc.tex
In the present paper, we have investigated the gene expression dynamics under noise, together with a positive feedback change for the epigenetic factors. Through the evolutionary change, we have succeeded in obtaining GRNs that are capable of differentiation to generate multiple cell types. The dynamics for differentiation are identified, categorized into three basic types (A, B, C), and analyzed in terms of dynamical systems, as summarized below.\\
\\
Here, we have three types of differentiation processes in our epigenetic modification model. Since the classification is based on initial attractors, we can confidently state that types A and C account for all the well-established attractors in dynamical systems theory (chaos, limit cycles, and fixed points). For each type, the later time course of differentiation is identified as in Table \ref{Tabel_ref}. Through extensive exploration of the examples that generate many cell types, they are categorized into the aforementioned three types. However, since it is not possible to rule out other possibilities of time courses empirically, we therefore focus our discussion on the three observed types without ruling out that others may arise.\\
\\
Type A adopts chaotic gene-expression dynamics, which are replaced by a few periodic attractors, representing oscillatory gene expressions. Then, these oscillatory dynamics are replaced gradually through the change in epigenetic modification towards several stable fixed points representing distinct cell types. Here, initial chaotic dynamics amplify cell-to-cell differences in gene expression, as a result of orbital instability. Then they are differentiated to subcycles which are later stabilized into fixed points corresponding to the  epigenetic changes.\\
\\
This is reminiscent of the previously proposed hierarchical attractor generation from limit-cycle\cite{Matsushita_2020}, whereas here, the initial attractor before the start of epigenetic modification is chaotic. By chaotic dynamics, gene-expression states are diversified by cells, which allows for the generation of multiple states, organized hierarchically. (See also\cite{suzuki2011oscillatory,koseska2013transition} and\cite{furusawa2001theory,furusawa2009chaotic,huang2005cell} for the possible role of oscillation and chaotic gene expression to pluripotency, respectively.)\\
\\
In type B, the gene expression dynamics are restricted within a low (say two) dimensional space, within which fixed-point attractors continuously exist depending on the epigenetic modification levels. For these states, the expression levels of a few genes take intermediate values between on and off, and these fixed points are easily moved by noise. Then, under the epigenetic modification change, the fixed points migrate to a few directions so that the stability of the fixed points is increased (i.e., so that they are less vulnerable to noise) until the expressions finally fixate to on or off. Here, the state changes occur under low-dimensional channels corresponding to the slow changes in the expression of the few genes. The cellular state changes slowly within the channel, driven by the noise and fixated by the epigenetic change, towards stable differentiated states. \\
\\
Here, noise and epigenetic modifications are essential to drive differentiation, as well as the presence of a few genes that slowly change their values (towards 1 or -1). Such changes act as a driver of the differentiation into a few given cell types, where the expression change induced by the noise increases the stability through epigenetic modifications. It is of interest that although the noise is random and causes no directional motion per se, the epigenetic fixation process gives rise to directionality. The slow positive feedback with the change in $\theta_i$ results in directional motion in gene expression levels $x_i$, as observed in the ratchet mechanism\cite{himeoka2020epigenetic}. \\
\\
In the type C, multiple fixed points are generated at an initial stage, which are separated by saddle points. Due to noise, different fixed points are reached and are further fixated by epigenetic changes. This differentiation process induces an early cell fate decision process from spread orbits. Due to this, the cell fates are heavily dependent on their initial conditions. Hence, this process lacks robustness in the cell-type distribution. \\
\\
%\textcolor{red}{The classification of the types based on their initial attractor tends to be sufficient, as most differentiation mechanisms occur early in the evolved networks. Nevertheless, there are still exceptions to this rule, and some nuance is required when considering the later time course. Combination types are one such example, where an initial type C saddle point differentiation does not lead to a fixed point, but rather to a secondary differentiation mechanism. Additionally, these networks have on some occasions exhibited transitions in their behaviour, such as going from a type A or C to a type B. We have excluded these networks from our analysis and do not regard them as an additional type. Instead, one could consider them a series of types A, B, or C differentiations happening in sequence.}\\
%\\
\begin{figure*}
    \includegraphics[width=1\linewidth]{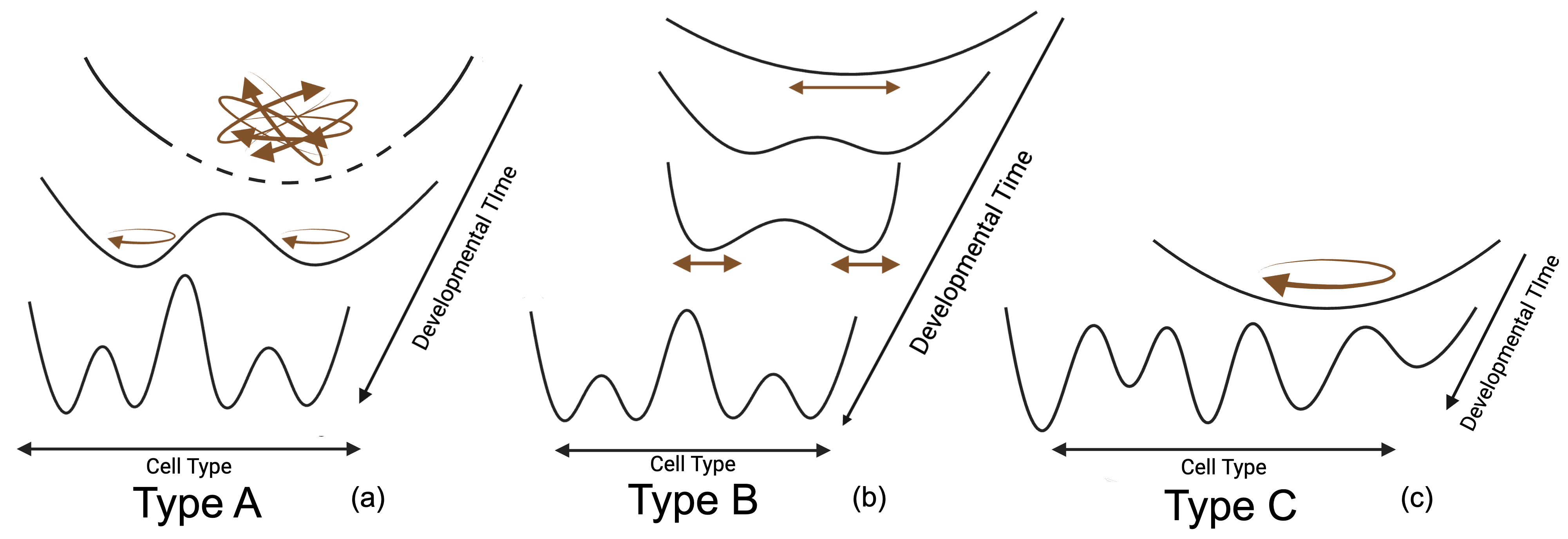}
    
    \caption{Simplified schematic of the differentiation process for Waddington's landscape. Type A undergoes initial chaotic dynamics, transitions to (quasi-)periodic cycles, and finally falls into fixed points. For type B, there are no cycles present, but rather, the landscape shifts to determine where the stable fixed point is located. Cells can end on either side of the initial potential hills through noise however, this process becomes irreversible as cell fate commitment progresses. Type C networks do not slowly transition into their final cell states but instead have a rather abrupt transition as the dynamics suddenly quench.}
    \label{fig:SchematicPlot}
\end{figure*}
Since $J_{ij}$ is not symmetric, the gene expression dynamics cannot be generally represented by a flow that falls down along a potential gradient. 
%Furthermore, stable regions for gene expression dynamics are determined by the epigenetic factors $\theta$, which differ between cells. Due to this fixation, each cell converts its own local minima into its global minima for the gene expression phase space $x_i$. Other minima can be found by adjusting the epigenetic factors, which change the position and behavior of the global attractor state. 
Still, in Morse-Smale systems\cite{smale1961gradient,rand2021geometry}, this global attraction can be represented by a potential except for the dynamics in the vicinity of the attractor. Generally speaking, bounded dynamical systems with a finite number of limit cycles and fixed point attractors fall under Morse-Smale systems; thus, types B and C can initially be represented by a potential landscape. In contrast, type A exhibits chaotic attractors that do not satisfy the conditions for Morse-Smale and thus a potential is not guaranteed mathematically; however, since there is global attraction in a bounded system, we may still draw a schematic representation of the landscape. Additionally, the usage of a potential
landscape as visualized by Waddington is still insightful for visualizing and understanding the generation of distinct orbits and fixed points generated from the initial
chaotic orbit. Here, with the slow change in $\theta$, gene expression dynamics for $x_i$ are changed. Hence, the landscape slowly changes, as represented by the changes along the depth axis in Waddington's picture.\\
\\
Following the argument above, the three process for cell differentiation are schematically summarized in Fig. \protect\ref{fig:SchematicPlot}, inspired by Waddington’s landscape. Types A and B exhibit the formation of successive valleys as proposed by Waddington. In type A, a shallow valley with chaotic dynamics splits into smaller valleys that correspond to different periodic attractors, which are then replaced by deeper valleys corresponding to fixed points in dynamical systems. This globally attracting chaotic attractor may schematically be represented by a process falling along a valley. The later differentiation of cycles or fixed points is represented by branching valleys.
The landscape as in Fig. \protect\ref{fig:SchematicPlot}.a may represent this. However, since chaotic dynamics are not covered by a Morse-Smale system, future studies will be necessary.\\
\\
In type B, a shallow valley corresponding to weakly stable fixed points is moved to two distinct directions and gradually deepens, corresponding to the increase in stability, as a result of the migration of fixed points and epigenetic fixation. This differentiation process also comes closer to fulfilling the conditions for a Morse-Smale system. Here, the epigenetic fixation and increase in stability can be represented by a parametrization of the landscape depending on $\theta$. A cell state settles into a fixed point, a stochastic perturbation occurs and updates the parametrization of the landscape according to the change in $\theta$ from this perturbation,
moving the fixed point slightly. Alternatively, one can define a continuous representation of the parametrized landscape by also using the epigenetic factor $\theta$ as a depth coordinate. The gradual increase in the stabilization of type B may fit Waddington's landscape. (see Fig. \ref{fig:SchematicPlot}.b)

As the gene expression dynamics in type C fall on the Morse-Smale system, the potential landscape can be generated. The behavior of the quenching fixed points through unstable orbits and saddle points fit well and can be represented by a Mores-Smale system, similarly the quenching periodic attractor can be obtained through SNIC bifurcations as visualized in Fig. 1.G in Rand et al. \cite{rand2021geometry}. However multiple fixed points become attractors all at once. Hence, multiple deep valleys are generated almost simultaneously within a short time span. This rapid fixation may deviate from Waddington’s picture since cell fate decision happens suddenly rather than through hierarchic splitting of valleys. (see Fig. \ref{fig:SchematicPlot}.c)\\
\\
Next, the robustness to perturbation to noise and variation in initial conditions is achieved in types A and B but not in type C. In type C, the resultant cell types may crucially depend on the initial condition or samples with different noise realizations. Even if the same cell types are generated by samples, the number distribution of each cell type depends crucially on each run of simulations and its initial conditions. Hence, type C would not be appropriate to be adopted in cell systems. In fact, type C is mostly evolved when the noise level in the expression dynamics is low, whereas the gene expression dynamics are rather noisy. Given the postulate that the differentiation works under larger stochasticity, it would be reasonable to assume types A and B to be more appropriate. \\
\\
%Increases in variance for complex systems act as an indicator for transitions\cite{scheffer2012anticipating}. 
High variations and fluctuations in gene expression levels play an important role in the differentiation process and are thought to be induced by noise\cite{eldar2010functional}. In our work, the increase in variance is observed in Fig. \ref{fig:TimeCourseVariances} during the differentiation process. Interestingly, type B networks exhibit a decrease of variance in the middle stage during differentiation as the differentiation process adopts stabilization of the gene expression levels after an initial dispersion.\\
\\
The variance of gene expression levels in cell populations generally increases due to the varying cell fates and heterogeneous pluripotent states that have been observed\cite{canham2010functional}. Cells with shared final cell fates exhibit a peak in their variance that reduces after differentiation\cite{gao2023single,noller2024cell}. This intra-cell variance,  shown in Fig. \ref{Fig:Intrapheno} indicates how gene expression states with large variation are eventually stabilized to the same final cell fate. Gene expression dynamics need to show a high degree of plasticity to comply with the high increase in variance for cellular populations and the strong decrease in intra-cell variance leading to stable final cell fates.\\
\\
The migration of fixed points with weak stability in type B suggests the emergence of slow expression change of a few genes. Slow changes in the expression levels stabilize the cellular states, which also control and stabilize the expressions of other genes. Such low-dimensional slow modes have been noted in the numerical evolution of morphogenesis by reaction-diffusion process\cite{kohsokabe2022dynamical}, as well as in the numerical evolution of intracellular reaction dynamics\cite{furusawa2018formation, sato2020evolutionary}. In these cases, the emergence of a few slow modes leads to the dimensional reduction of phenotypes, which can increase the controllability of the gene-expression states and evolvability\cite{kaneko2024constructing}. The relevance of such a slowly changing expression to the cell differentiation process and morphogenesis, as well as the generality of dimensional reduction, needs to be explored further in both simulations and experiments.\\
\\
In the type B case, fixed-point attractors for gene expressions $x_i$ are located on a line depending on the change in epigenetic modification $\theta_i$. This may remind one of the line-attractors studied extensively in neural networks\cite{mante2013context,seung1996brain}. In the present case, however, the fixed points are not for identical dynamical systems sharing $\theta_i$'s but are for dynamical systems with different $\theta_i$. In both cases, fixed points are strongly attracted from the directions orthogonal to the lines, leading to the motion along the line to be more feasible. Such separation of scales between on and off manifolds will also be relevant to control in gene expression dynamics (or neural systems as well).
Note that the dynamics with the on$/$off states with a step-like function as in the present model, are also adopted by neural networks and spin-glass problems in statistical physics. In the latter, the emergence of multi-stable states against quenching randomness has been explored (as in type C), whereas annealing that slowly changes the connection matrix $J_{ij}$ prunes some metastable states. The existence of annealing and quenching, observed for types B and C, is of interest in this regard, although here the change occurs not by the change in $J_{ij}$ but with the autonomous change in $\theta_i$. Extension of the statistical-physics approach to the present case will be of interest in the future.\\
\\
Furthermore, a natural extension of the model would be to include spatial effects such as through reaction-diffusion\cite{green2015positional}, lateral inhibition\cite{meinhardt2000pattern}, or morphogen gradient signaling\cite{rogers2011morphogen}. While the cells in our model are not spatially configured, biological development involves spatial organization which requires positional information for differentiation \cite{wolpert1969positional}. A simple way to extend the current model is through the inclusion of morphogen gradients as inputs to the GRN. Cells can be arranged on a 1D line or a 2D square and produce robust spatial patterns through the information provided by the morphogen. Previous works have discussed the information that can be provided by noisy morphogens\cite{tkavcik2021many} and how such systems may be optimized for self-organisation\cite{bruckner2024information}. Spatio-temporal averaging through the epigenetic factor $\theta$ may also play an important role in circumventing errors induced by noise or reading out many bits of information through temporally averaging morphogen gradients. Additionally, the model would be ideal in understanding how morphogen signals are translated to the cell fate decisions and leads to canalization through GRNs. Our preliminary results have already shown the type B process to be able to extract multiple bits of information from a single noisy morphogen through gene concentration in intermediate values as discussed by Tkačik and Gregor \cite{tkavcik2021many}. \\
\\
Lateral inhibition as done through Notch signalling, would also be a valuable extension for the current model. Here, bistable states that undergo developmental patterning are created in nondividing cells\cite{corson2017self} and during regeneration\cite{schwayer2025cell}. These behaviors are reminiscent of type B's weakly stable fixed points, and lateral inhibition may be an alternative process to induce directed diffusion leading to differentiation. Future studies are required to investigate how type B will behave with similar lateral inhibition models.\\
\\

%% file: 5.Disc/1ExpRel.tex
Finally, we discuss possible connections between the present study and experimental studies.
Through the abundance of mRNA molecules extrapolated from dissociated cells, it has been possible to obtain data on the gene expression patterns through the course of cell differentiation. Techniques to trace cell fate commitments that reconstruct the cellular lineage and its gene expression levels have gone through major advancements. Nevertheless, there are several limitations to tracking the gene expression dynamics at a single cell level and tracking the epigenetic modification levels, which is much more difficult.\\
\\
Noting such limitations in mind, we here discuss a possible connection between our theoretical results with experimental results.
For example, the transcriptional uncertainty in stem cells is examined by Gao et al.\cite{gao2023single} by computing the negative log-likelihood of an individual cell's gene expression levels with respect to the gene expressions of an interpolated cell-fate decision process. They reconstructed the cell-to-cell variability detected in other differentiation processes\cite{dussiau2022hematopoietic,richard2016single}. The phenotypic variance, calculated in Fig. \ref{Fig:Intrapheno}, is comparable to their log-likelihood as they both measure the variation between cells of a given cell type. Both the experimental data and our model show a salient peak of the variance before cell fate commitment. Cells undergo this high variance state as an exploratory phase to ensure that all the possible final states can be reached while the variance is reduced as a result of the commitment and stabilization of the final cell type. The results for types A and B suggest a gradual decrease in the variance after the peak, which can be verified experimentally in the future.\\
\\
Another example of experimental comparison is the reduction in the dimensionality of gene expressions analyzed by Biondo et al.\cite{Biondo_2024}, where the intrinsic dimension of transcriptomic data is measured and used as a proxy for the differentiation potential of a cell in Waddington's landscape. Cell type potency is reduced with the decrease in intrinsic dimension throughout differentiation. Such dimensional reduction is consistent with our results for types A and B. \\
\\
Other measures could be adopted in experimental data and theoretical models to deepen our understanding of the process behind differentiation. For example, in the experimental analysis of cellular robustness under perturbation, the presence and fraction of genes whose expression levels fluctuate, oscillate, or slowly change will be important. For instance, Moussy et al.\cite{Moussy_2017} tracked transcriptional and morphological changes in hematopoietic cells during cell fate commitment and found that cells can exhibit intermediate states rather than a simple binary switch. Oscillations in the methylation of genomes have also been observed in experiments\cite{kangaspeska2008transient,rulands2018genome}, supporting the idea of differentiation through oscillatory gene expression dynamics\cite{matsushita2022dynamical}.  \\
\\
So far, it is difficult to judge whether type A or type B is more appropriate for experimentally observed cell differentiations. Both exhibit a Waddington-type epigenetic landscape, robust irreversible cell differentiation under noise, dimensional reduction in expression dynamics, and transient increase in the variance. Some reports suggest the presence of oscillatory gene expression in the initial stages of development or in pluripotent cells\cite{kobayashi2009cyclic,imayoshi2013oscillatory,aulehla2008oscillating}, whereas the cell-cell variation by stochastic gene expression has been noted in other studies\cite{singer2014dynamic,elowitz2002stochastic}.\\
\\
In conclusion, we have identified and analyzed three fundamental processes of cell differentiation through evolutionary simulations of gene expression dynamics with a slow epigenetic fixation process: (a) successive fixation of chaotic dynamics, (b) annealing of fixed points via noise and epigenetic fixation, and (c) quenching of fixed points. The first two processes exhibit robustness against initial perturbations and noise, generating an epigenetic landscape consistent with Waddington’s classical picture. Our findings provide new insights into the dynamical systems framework for cell differentiation, incorporating slow/fast time scales that align with experimental observations and pave the way for further exploration of cell differentiation from an evolutionary and dynamical systems perspective.

%% file: Appendix.tex
\begin{figure*}[htbp]
    \centering
    \includegraphics[width=1\linewidth]{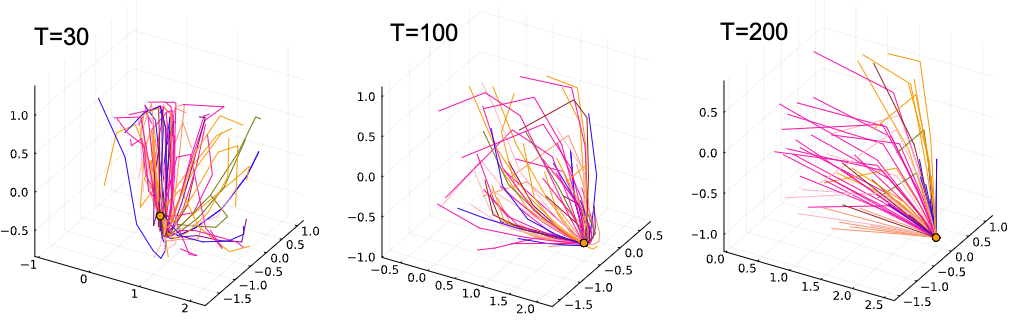}
    \caption{{Noiseless orbits of cells for type B network by giving the same $\theta_i$ values. At the indicated times, the cells $\theta_i$ values are manually set to those of the first cell. The cells that are spread apart suddenly converge onto the fixed point of the first cell (indicated by the circle), demonstrating the presence of only a single fixed point and that its location is fully determined by its epigenetic factors $\theta_i$ rather than its gene expression $x_i$. We have performed this analysis at various times, with $\theta_i$ of different cells and multiple type B networks, all with the same results.}}
    \label{fig:BSingleEpi}
\end{figure*}

To confirm this single fixed point attractor for given $\theta$ for type B, we plotted several orbits for $x_i$, starting from different initial conditions in Fig. \ref{fig:BSingleEpi}. We use the positions of the 64 cells to generate the different initial conditions. This allows us to sample the system's behavior in the subspace where the dynamics of the gene expression levels of our cells reside. While we cannot exclude that other fixed points exist for highly perturbed systems, these are either too distant and cannot be reached by the network's dynamics, or their basin of attraction is too small to play any role. By setting $\theta$ for all cells equal to those at a given developmental time $T$, a singular, distinct fixed point is reached in each case. All cells converge to this identical fixed point, indicating the absence of other fixed points for type B networks and that the position of fixed points is fully determined by the epigenetic modification level $\theta_i$. \\
\\
\begin{figure}[!h]
    \centering
    \includegraphics[width=1\linewidth]{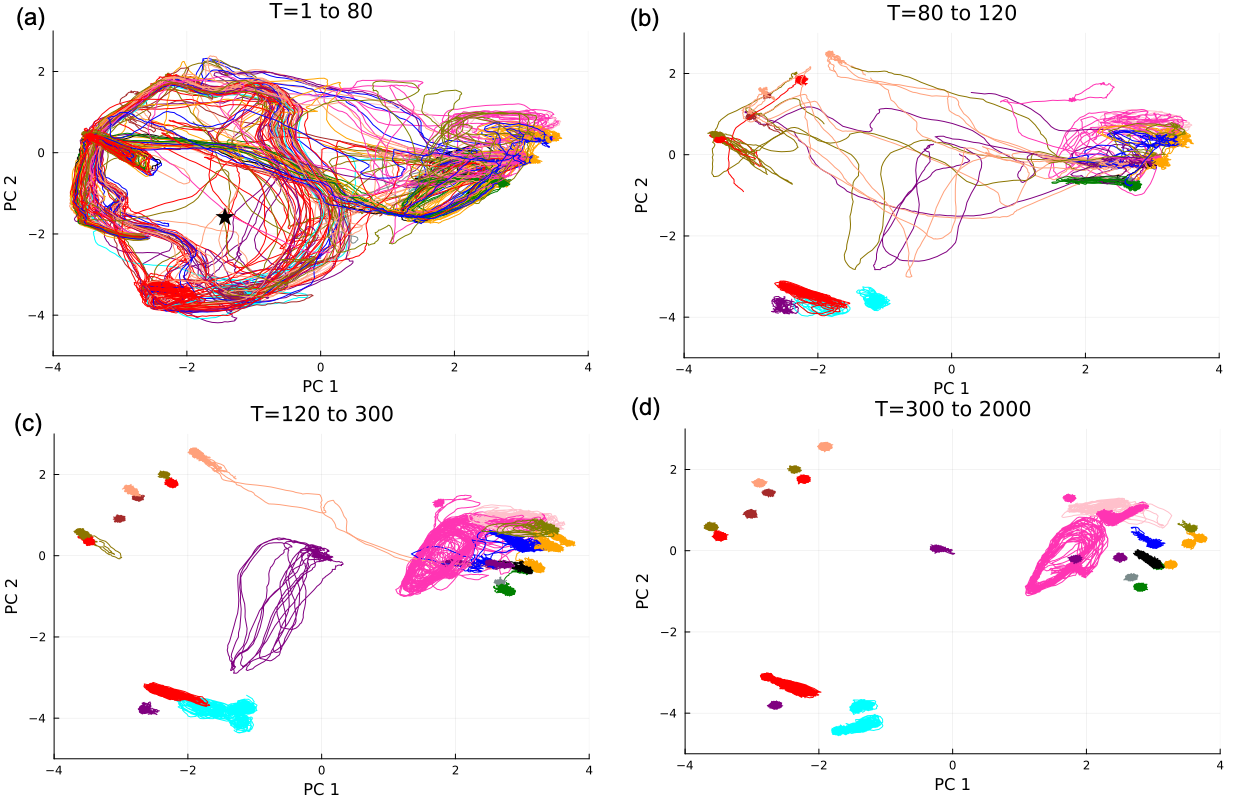}
    \caption{2D Orbits in PC space at various stages of differentiation for combination type. The star on the first plot indicates the origin, and color is used to indicate the different final cell types. For these gene dynamics, we observe type C in the first plot and types A and B in the next 3 plots.}
    \label{fig:Combi_Oribts}
\end{figure}\\
\begin{figure}[!h]
    \centering
    \includegraphics[width=1\linewidth]{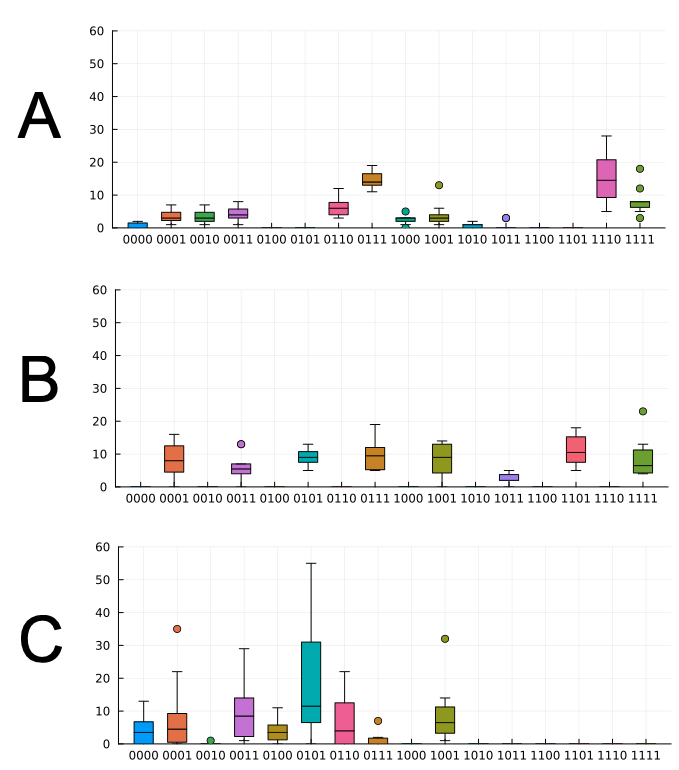}
    
    \caption{Target cell type distribution of a single typical A, B, and C network for 10 simulations with perturbed initial conditions (radius of 5, $\pm 0.79$ per $x_i$). The amount of a given target cell (y-axis) remains relatively constant for A and B; for C, it can fluctuate from 0 to almost 90\% of all cells.}
    \label{fig:Boxplot}
\end{figure}\\
While we have already discussed idealized cases, differentiation in some other networks consists of combinations of the three processes. We show one such network in Fig. \ref{fig:Combi_Oribts} that has all 3 previously mentioned processes throughout differentiation.\\
\\
The initial periodic attractor, located at PC 1 $< 0.5$, is quickly reached (Fig. \ref{fig:Combi_Oribts}.a). This global attractor quickly disappears and is quenched rather than evolving or migrating into a new shape. This can be observed through the cells with various fates that migrate from (-3,0) towards a cycle on the right at (1,-2). At $T=80$ (Fig. \ref{fig:Combi_Oribts}.b), none of the cells are present in this initial attractor, rather they have formed 3 smaller clusters with their own secondary differentiation process. Fig. \ref{fig:Combi_Oribts}.c demonstrates these branched cell dynamics, in the top left corner, cells end up falling on fixed points which slowly move and branch into 2 separate final cell states. This branching of weakly stable fixed points and epigenetic stabilization is akin to the type B process. Cells at the top right corner fall into periodic attractors that change shape, disappear, and eventually become fixed points as observed for type A (Fig. \ref{fig:Combi_Oribts}.d). In the bottom left corner is another chaotic attractor from which the differentiation progresses \\
\\
In Fig. \ref{fig:Boxplot}, we plot the number of each cell type when perturbing the initial conditions. For simplicity, we only plot a single network for each of the three types, further analysis has been performed for all samples used in the robustness plot Fig. \ref{fig:Perturb_Rob}. For types A and B, the number of each cell type is mostly unaffected by the perturbations with their half-width below 10 cells (except for cell type 1110 for type A) and relatively few, small outliers. Type C, however, has 3 target cell types with a half-width above 10 and 3 target cell types whose full-width or outliers have a range of over half the total count of 64 cells.\\ 
\\

%% file: apssamp.bib
@PREAMBLE{
 "\providecommand{\noopsort}[1]{}" 
 # "\providecommand{\singleletter}[1]{#1}%" 
}

@book{waddington1957strategy,
  title={The Strategy of the Genes: A Discussion of Some Aspects of Theoretical Biology},
  author={Waddington, C.H.},
  lccn={lc58001976},
  url={https://books.google.co.uk/books?id=PdU9AAAAIAAJ},
  year={1957},
  publisher={Allen \& Unwin}
}

@article{smale1961gradient,
  title={On gradient dynamical systems},
  author={Smale, Stephen},
  journal={Annals of Mathematics},
  volume={74},
  number={1},
  pages={199--206},
  year={1961},
  publisher={JSTOR}
}

@article{rand2021geometry,
  title={Geometry of gene regulatory dynamics},
  author={Rand, David A and Raju, Archishman and S{\'a}ez, Meritxell and Corson, Francis and Siggia, Eric D},
  journal={Proceedings of the National Academy of Sciences},
  volume={118},
  number={38},
  pages={e2109729118},
  year={2021},
  publisher={National Academy of Sciences}
}

@article{meinhardt2000pattern,
  title={Pattern formation by local self-activation and lateral inhibition},
  author={Meinhardt, Hans and Gierer, Alfred},
  journal={Bioessays},
  volume={22},
  number={8},
  pages={753--760},
  year={2000},
  publisher={Wiley Online Library}
}

@article{corson2017self,
  title={Self-organized Notch dynamics generate stereotyped sensory organ patterns in Drosophila},
  author={Corson, Francis and Couturier, Lydie and Rouault, Herv{\'e} and Mazouni, Khalil and Schweisguth, Fran{\c{c}}ois},
  journal={Science},
  volume={356},
  number={6337},
  pages={eaai7407},
  year={2017},
  publisher={American Association for the Advancement of Science}
}

@article{schwayer2025cell,
  title={Cell heterogeneity and fate bistability drive tissue patterning during intestinal regeneration},
  author={Schwayer, C and Barbiero, S and Br{\"u}ckner, DB and Baader, C and Repina, NA and Diaz, OE and Meylan, L Challet and Kalck, V and Suppinger, S and Yang, Q and others},
  journal={bioRxiv},
  pages={2025--01},
  year={2025},
  publisher={Cold Spring Harbor Laboratory}
}

@article{bruckner2024information,
  title={Information content and optimization of self-organized developmental systems},
  author={Br{\"u}ckner, David B and Tka{\v{c}}ik, Ga{\v{s}}per},
  journal={Proceedings of the National Academy of Sciences},
  volume={121},
  number={23},
  pages={e2322326121},
  year={2024},
  publisher={National Academy of Sciences}
}

@article{tkavcik2021many,
  title={The many bits of positional information},
  author={Tka{\v{c}}ik, Ga{\v{s}}per and Gregor, Thomas},
  journal={Development},
  volume={148},
  number={2},
  pages={dev176065},
  year={2021},
  publisher={The Company of Biologists Ltd}
}

@Article{Coomer2022,
author={Coomer, Megan A.
and Ham, Lucy
and Stumpf, Michael P.H.},
title={Noise distorts the epigenetic landscape and shapes cell-fate decisions},
journal={Cell Systems},
year={2022},
month={Jan},
day={19},
publisher={Elsevier},
volume={13},
number={1},
pages={83-102.e6},
issn={2405-4712},
doi={10.1016/j.cels.2021.09.002},
url={https://doi.org/10.1016/j.cels.2021.09.002}
}

@article{Furusawa2012,
  title={A dynamical-systems view of stem cell biology},
  author={Furusawa, Chikara and Kaneko, Kunihiko},
  journal={Science},
  volume={338},
  number={6104},
  pages={215--217},
  year={2012},
  publisher={American Association for the Advancement of Science}
}

@article{mante2013context,
  title={Context-dependent computation by recurrent dynamics in prefrontal cortex},
  author={Mante, Valerio and Sussillo, David and Shenoy, Krishna V and Newsome, William T},
  journal={nature},
  volume={503},
  number={7474},
  pages={78--84},
  year={2013},
  publisher={Nature Publishing Group UK London}
}

@article{elowitz2002stochastic,
  title={Stochastic gene expression in a single cell},
  author={Elowitz, Michael B and Levine, Arnold J and Siggia, Eric D and Swain, Peter S},
  journal={Science},
  volume={297},
  number={5584},
  pages={1183--1186},
  year={2002},
  publisher={American Association for the Advancement of Science}
}

@article{vinuelas2013quantifying,
  title={Quantifying the contribution of chromatin dynamics to stochastic gene expression reveals long, locus-dependent periods between transcriptional bursts},
  author={Vi{\~n}uelas, Jos{\'e} and Kaneko, Ga{\"e}l and Coulon, Antoine and Vallin, Elodie and Morin, Val{\'e}rie and Mejia-Pous, Camila and Kupiec, Jean-Jacques and Beslon, Guillaume and Gandrillon, Olivier},
  journal={BMC biology},
  volume={11},
  pages={1--19},
  year={2013},
  publisher={Springer}
}

@article{mcadams1997stochastic,
  title={Stochastic mechanisms in gene expression},
  author={McAdams, Harley H and Arkin, Adam},
  journal={Proceedings of the National Academy of Sciences},
  volume={94},
  number={3},
  pages={814--819},
  year={1997},
  publisher={The National Academy of Sciences of the USA}
}

@article{furusawa2005ubiquity,
  title={Ubiquity of log-normal distributions in intra-cellular reaction dynamics},
  author={Furusawa, Chikara and Suzuki, Takao and Kashiwagi, Akiko and Yomo, Tetsuya and Kaneko, Kunihiko},
  journal={Biophysics},
  volume={1},
  pages={25--31},
  year={2005},
  publisher={The Biophysical Society of Japan}
}

@article{huang2005cell,
  title={Cell fates as high-dimensional attractor states of a complex gene regulatory network},
  author={Huang, Sui and Eichler, Gabriel and Bar-Yam, Yaneer and Ingber, Donald E},
  journal={Physical review letters},
  volume={94},
  number={12},
  pages={128701},
  year={2005},
  publisher={APS}
}

@article{aulehla2008oscillating,
  title={Oscillating signaling pathways during embryonic development},
  author={Aulehla, Alexander and Pourqui{\'e}, Olivier},
  journal={Current opinion in cell biology},
  volume={20},
  number={6},
  pages={632--637},
  year={2008},
  publisher={Elsevier}
}

@article{villani2011dynamical,
  title={A dynamical model of genetic networks for cell differentiation},
  author={Villani, Marco and Barbieri, Alessia and Serra, Roberto},
  journal={PloS one},
  volume={6},
  number={3},
  pages={e17703},
  year={2011},
  publisher={Public Library of Science San Francisco, USA}
}

@article{wang2010potential,
  title={The potential landscape of genetic circuits imposes the arrow of time in stem cell differentiation},
  author={Wang, Jin and Xu, Li and Wang, Erkang and Huang, Sui},
  journal={Biophysical journal},
  volume={99},
  number={1},
  pages={29--39},
  year={2010},
  publisher={Elsevier}
}

@article{brackston2018transition,
  title={Transition state characteristics during cell differentiation},
  author={Brackston, Rowan D and Lakatos, Eszter and Stumpf, Michael PH},
  journal={PLoS computational biology},
  volume={14},
  number={9},
  pages={e1006405},
  year={2018},
  publisher={Public Library of Science San Francisco, CA USA}
}

@article{newman2020cell,
  title={Cell differentiation: What have we learned in 50 years?},
  author={Newman, Stuart A},
  journal={Journal of theoretical biology},
  volume={485},
  pages={110031},
  year={2020},
  publisher={Elsevier}
}

@article{jaeger2014bioattractors,
  title={Bioattractors: dynamical systems theory and the evolution of regulatory processes},
  author={Jaeger, Johannes and Monk, Nick},
  journal={The Journal of physiology},
  volume={592},
  number={11},
  pages={2267--2281},
  year={2014},
  publisher={Wiley Online Library}
}

@article{green2015positional,
  title={Positional information and reaction-diffusion: two big ideas in developmental biology combine},
  author={Green, Jeremy and Sharpe, James},
  journal={Development},
  volume={142},
  number={7},
  pages={1203--1211},
  year={2015},
  publisher={The Company of Biologists}
}

@article{rogers2011morphogen,
  title={Morphogen gradients: from generation to interpretation},
  author={Rogers, Katherine W and Schier, Alexander F},
  journal={Annual review of cell and developmental biology},
  volume={27},
  number={1},
  pages={377--407},
  year={2011},
  publisher={Annual Reviews}
}

@article{wolpert1969positional,
  title={Positional information and the spatial pattern of cellular differentiation},
  author={Wolpert, Lewis},
  journal={Journal of theoretical biology},
  volume={25},
  number={1},
  pages={1--47},
  year={1969},
  publisher={Elsevier}
}

@article{singer2014dynamic,
  title={Dynamic heterogeneity and DNA methylation in embryonic stem cells},
  author={Singer, Zakary S and Yong, John and Tischler, Julia and Hackett, Jamie A and Altinok, Alphan and Surani, M Azim and Cai, Long and Elowitz, Michael B},
  journal={Molecular cell},
  volume={55},
  number={2},
  pages={319--331},
  year={2014},
  publisher={Elsevier}
}

@article{kaneko2024constructing,
  title={Constructing universal phenomenology for biological cellular systems: an idiosyncratic review on evolutionary dimensional reduction},
  author={Kaneko, Kunihiko},
  journal={Journal of Statistical Mechanics: Theory and Experiment},
  volume={2024},
  number={2},
  pages={024002},
  year={2024},
  publisher={IOP Publishing}
}

@article{rulands2018genome,
  title={Genome-scale oscillations in DNA methylation during exit from pluripotency},
  author={Rulands, Steffen and Lee, Heather J and Clark, Stephen J and Angermueller, Christof and Smallwood, S{\'e}bastien A and Krueger, Felix and Mohammed, Hisham and Dean, Wendy and Nichols, Jennifer and Rugg-Gunn, Peter and others},
  journal={Cell systems},
  volume={7},
  number={1},
  pages={63--76},
  year={2018},
  publisher={Elsevier}
}

@article{kangaspeska2008transient,
  title={Transient cyclical methylation of promoter DNA},
  author={Kangaspeska, Sara and Stride, Brenda and M{\'e}tivier, Rapha{\"e}l and Polycarpou-Schwarz, Maria and Ibberson, David and Carmouche, Richard Paul and Benes, Vladimir and Gannon, Frank and Reid, George},
  journal={Nature},
  volume={452},
  number={7183},
  pages={112--115},
  year={2008},
  publisher={Nature Publishing Group UK London}
}

@article{atlasi2017interplay,
  title={The interplay of epigenetic marks during stem cell differentiation and development},
  author={Atlasi, Yaser and Stunnenberg, Hendrik G},
  journal={Nature Reviews Genetics},
  volume={18},
  number={11},
  pages={643--658},
  year={2017},
  publisher={Nature Publishing Group UK London}
}

@article{furusawa2009chaotic,
  title={Chaotic expression dynamics implies pluripotency: when theory and experiment meet},
  author={Furusawa, Chikara and Kaneko, Kunihiko},
  journal={Biology direct},
  volume={4},
  pages={1--11},
  year={2009},
  publisher={Springer}
}

@article{canham2010functional,
  title={Functional heterogeneity of embryonic stem cells revealed through translational amplification of an early endodermal transcript},
  author={Canham, Maurice A and Sharov, Alexei A and Ko, Minoru SH and Brickman, Joshua M},
  journal={PLoS biology},
  volume={8},
  number={5},
  pages={e1000379},
  year={2010},
  publisher={Public Library of Science San Francisco, USA}
}

@article{furusawa2001theory,
  title={Theory of robustness of irreversible differentiation in a stem cell system: chaos hypothesis},
  author={Furusawa, Chikara and Kaneko, Kunihiko},
  journal={Journal of Theoretical Biology},
  volume={209},
  number={4},
  pages={395--416},
  year={2001},
  publisher={Elsevier}
}

@article{koseska2013transition,
  title={Transition from amplitude to oscillation death via Turing bifurcation},
  author={Koseska, Aneta and Volkov, Evgenii and Kurths, J{\"u}rgen},
  journal={Physical review letters},
  volume={111},
  number={2},
  pages={024103},
  year={2013},
  publisher={APS}
}

@article{suzuki2011oscillatory,
  title={Oscillatory protein expression dynamics endows stem cells with robust differentiation potential},
  author={Suzuki, Narito and Furusawa, Chikara and Kaneko, Kunihiko},
  journal={PloS one},
  volume={6},
  number={11},
  pages={e27232},
  year={2011},
  publisher={Public Library of Science San Francisco, USA}
}

@article{kaern2005stochasticity,
  title={Stochasticity in gene expression: from theories to phenotypes},
  author={Kaern, Mads and Elston, Timothy C and Blake, William J and Collins, James J},
  journal={Nature Reviews Genetics},
  volume={6},
  number={6},
  pages={451--464},
  year={2005},
  publisher={Nature Publishing Group UK London}
}

@article{raj2008nature,
  title={Nature, nurture, or chance: stochastic gene expression and its consequences},
  author={Raj, Arjun and Van Oudenaarden, Alexander},
  journal={Cell},
  volume={135},
  number={2},
  pages={216--226},
  year={2008},
  publisher={Elsevier}
}

@article{seung1996brain,
  title={How the brain keeps the eyes still},
  author={Seung, H Sebastian},
  journal={Proceedings of the National Academy of Sciences},
  volume={93},
  number={23},
  pages={13339--13344},
  year={1996},
  publisher={The National Academy of Sciences of the USA}
}

@article{Matsushita_2020,
   title={Homeorhesis in Waddington’s landscape by epigenetic feedback regulation},
   volume={2},
   ISSN={2643-1564},
   url={http://dx.doi.org/10.1103/PhysRevResearch.2.023083},
   DOI={10.1103/physrevresearch.2.023083},
   number={2},
   journal={Physical Review Research},
   publisher={American Physical Society (APS)},
   author={Matsushita, Yuuki and Kaneko, Kunihiko},
   year={2020},
   month=apr 
}

@article{waddington1942epigenotype,
  title={The epigenotype},
  author={Waddington, Conrad H},
  journal={Endeavour},
  volume={1},
  pages={18--20},
  year={1942}
}

@article{deichmann2016epigenetics,
  title={Epigenetics: The origins and evolution of a fashionable topic},
  author={Deichmann, Ute},
  journal={Developmental biology},
  volume={416},
  number={1},
  pages={249--254},
  year={2016},
  publisher={Elsevier}
}

@article{willbanks2016evolution,
  title={The evolution of epigenetics: from prokaryotes to humans and its biological consequences},
  author={Willbanks, Amber and Leary, Meghan and Greenshields, Molly and Tyminski, Camila and Heerboth, Sarah and Lapinska, Karolina and Haskins, Kathryn and Sarkar, Sibaji},
  journal={Genetics \& epigenetics},
  volume={8},
  pages={GEG--S31863},
  year={2016},
  publisher={SAGE Publications Sage UK: London, England}
}

@article{stillman2018histone,
  title={Histone modifications: insights into their influence on gene expression},
  author={Stillman, Bruce},
  journal={Cell},
  volume={175},
  number={1},
  pages={6--9},
  year={2018},
  publisher={Elsevier}
}

@article{matsushita2022dynamical,
  title={Dynamical systems theory of cellular reprogramming},
  author={Matsushita, Yuuki and Hatakeyama, Tetsuhiro S and Kaneko, Kunihiko},
  journal={Physical Review Research},
  volume={4},
  number={2},
  pages={L022008},
  year={2022},
  publisher={APS}
}

@article{mjolsness1991connectionist,
  title={A connectionist model of development},
  author={Mjolsness, Eric and Sharp, David H and Reinitz, John},
  journal={Journal of theoretical Biology},
  volume={152},
  number={4},
  pages={429--453},
  year={1991},
  publisher={Elsevier}
}

@article{gao2023single,
  title={Single-cell transcriptional uncertainty landscape of cell differentiation},
  author={Gao, Nan Papili and Gandrillon, Olivier and Pa\'{l}di, Andra\'{s} and Herbach, Ulysse and Gunawan, Rudiyanto},
  journal={F1000Research},
  volume={12},
  pages={426},
  year={2023}
}

@article{richard2016single,
  title={Single-cell-based analysis highlights a surge in cell-to-cell molecular variability preceding irreversible commitment in a differentiation process},
  author={Richard, Angélique and Boullu, Loïs and Herbach, Ulysse and Bonnafoux, Arnaud and Morin, Valérie and Vallin, Elodie and Guillemin, Anissa and Papili Gao, Nan and Gunawan, Rudiyanto and Cosette, Jérémie and others},
  journal={PLoS biology},
  volume={14},
  number={12},
  pages={e1002585},
  year={2016},
  publisher={Public Library of Science San Francisco, CA USA}
}

@article{dussiau2022hematopoietic,
  title={Hematopoietic differentiation is characterized by a transient peak of entropy at a single-cell level},
  author={Dussiau, Charles and Boussaroque, Agathe and Gaillard, Mathilde and Bravetti, Clotilde and Zaroili, Laila and Knosp, Camille and Friedrich, Chlo{\'e} and Asquier, Philippe and Willems, Lise and Quint, Laurent and others},
  journal={BMC biology},
  volume={20},
  number={1},
  pages={60},
  year={2022},
  publisher={Springer}
}

@article {Biondo_2024,
	author = {Biondo, Marta and Cirone, Niccol{\`o} and Valle, Filippo and Lazzardi, Silvia and Caselle, Michele and Osella, Matteo},
	title = {The intrinsic dimension of gene expression during cell differentiation},
	elocation-id = {2024.08.02.606382},
	year = {2024},
	doi = {10.1101/2024.08.02.606382},
	publisher = {Cold Spring Harbor Laboratory},
	


	journal = {bioRxiv}
}

@article{sneppen2008ultrasensitive,
  title={Ultrasensitive gene regulation by positive feedback loops in nucleosome modification},
  author={Sneppen, Kim and Micheelsen, Mille A and Dodd, Ian B},
  journal={Molecular systems biology},
  volume={4},
  number={1},
  pages={182},
  year={2008},
  publisher={John Wiley \& Sons, Ltd Chichester, UK}
}

@article{dodd2007theoretical,
  title={Theoretical analysis of epigenetic cell memory by nucleosome modification},
  author={Dodd, Ian B and Micheelsen, Mille A and Sneppen, Kim and Thon, Genevieve},
  journal={Cell},
  volume={129},
  number={4},
  pages={813--822},
  year={2007},
  publisher={Elsevier}
}

@article{kaneko2007evolution,
  title={Evolution of robustness to noise and mutation in gene expression dynamics},
  author={Kaneko, Kunihiko},
  journal={PLoS one},
  volume={2},
  number={5},
  pages={e434},
  year={2007},
  publisher={Public Library of Science San Francisco, USA}
}

@article{salazar2001phenotypic,
  title={Phenotypic and dynamical transitions in model genetic networks I. Emergence of patterns and genotype-phenotype relationships},
  author={Salazar-Ciudad, I and Newman, SA and Sol{\'e}, RV},
  journal={Evolution \& development},
  volume={3},
  number={2},
  pages={84--94},
  year={2001},
  publisher={Wiley Online Library}
}

@article{miyamoto2015pluripotency,
  title={Pluripotency, differentiation, and reprogramming: a gene expression dynamics model with epigenetic feedback regulation},
  author={Miyamoto, Tadashi and Furusawa, Chikara and Kaneko, Kunihiko},
  journal={PLoS computational biology},
  volume={11},
  number={8},
  pages={e1004476},
  year={2015},
  publisher={Public Library of Science San Francisco, CA USA}
}

@article{jacob1961genetic,
  title={Genetic regulatory mechanisms in the synthesis of proteins},
  author={Jacob, Fran{\c{c}}ois and Monod, Jacques},
  journal={Journal of molecular biology},
  volume={3},
  number={3},
  pages={318--356},
  year={1961},
  publisher={Elsevier}
}

@article{kauffman1969metabolic,
  title={Metabolic stability and epigenesis in randomly constructed genetic nets},
  author={Kauffman, Stuart A},
  journal={Journal of theoretical biology},
  volume={22},
  number={3},
  pages={437--467},
  year={1969},
  publisher={Elsevier}
}

@article{karlebach2008modelling,
  title={Modelling and analysis of gene regulatory networks},
  author={Karlebach, Guy and Shamir, Ron},
  journal={Nature reviews Molecular cell biology},
  volume={9},
  number={10},
  pages={770--780},
  year={2008},
  publisher={Nature Publishing Group UK London}
}

@article{schlitt2007current,
  title={Current approaches to gene regulatory network modelling},
  author={Schlitt, Thomas and Brazma, Alvis},
  journal={BMC bioinformatics},
  volume={8},
  pages={1--22},
  year={2007},
  publisher={Springer}
}

@article{Moussy_2017,
    author = {Moussy, Alice AND Cosette, Jérémie AND Parmentier, Romuald AND da Silva, Cindy AND Corre, Guillaume AND Richard, Angélique AND Gandrillon, Olivier AND Stockholm, Daniel AND Páldi, András},
    journal = {PLOS Biology},
    publisher = {Public Library of Science},
    title = {Integrated time-lapse and single-cell transcription studies highlight the variable and dynamic nature of human hematopoietic cell fate commitment},
    year = {2017},
    month = {07},
    volume = {15},
    url = {https://doi.org/10.1371/journal.pbio.2001867},
    pages = {1-23},
    number = {7},

}

@article{furusawa2018formation,
  title={Formation of dominant mode by evolution in biological systems},
  author={Furusawa, Chikara and Kaneko, Kunihiko},
  journal={Physical Review E},
  volume={97},
  number={4},
  pages={042410},
  year={2018},
  publisher={APS}
}

@article{sato2020evolutionary,
  title={Evolutionary dimension reduction in phenotypic space},
  author={Sato, Takuya U and Kaneko, Kunihiko},
  journal={Physical Review Research},
  volume={2},
  number={1},
  pages={013197},
  year={2020},
  publisher={APS}
}

@article{kobayashi2009cyclic,
  title={The cyclic gene Hes1 contributes to diverse differentiation responses of embryonic stem cells},
  author={Kobayashi, Taeko and Mizuno, Hiroaki and Imayoshi, Itaru and Furusawa, Chikara and Shirahige, Katsuhiko and Kageyama, Ryoichiro},
  journal={Genes \& development},
  volume={23},
  number={16},
  pages={1870--1875},
  year={2009},
  publisher={Cold Spring Harbor Lab}
}

@article{imayoshi2013oscillatory,
  title={Oscillatory control of factors determining multipotency and fate in mouse neural progenitors},
  author={Imayoshi, Itaru and Isomura, Akihiro and Harima, Yukiko and Kawaguchi, Kyogo and Kori, Hiroshi and Miyachi, Hitoshi and Fujiwara, Takahiro and Ishidate, Fumiyoshi and Kageyama, Ryoichiro},
  journal={Science},
  volume={342},
  number={6163},
  pages={1203--1208},
  year={2013},
  publisher={American Association for the Advancement of Science}
}

@article{himeoka2020epigenetic,
  title={Epigenetic Ratchet: Spontaneous adaptation via stochastic gene expression},
  author={Himeoka, Yusuke and Kaneko, Kunihiko},
  journal={Scientific Reports},
  volume={10},
  number={1},
  pages={459},
  year={2020},
  publisher={Nature Publishing Group UK London}
}

@article{kohsokabe2022dynamical,
  title={Dynamical systems approach to evolution--development congruence: Revisiting Haeckel's recapitulation theory},
  author={Kohsokabe, Takahiro and Kaneko, Kunihiko},
  journal={Journal of Experimental Zoology Part B: Molecular and Developmental Evolution},
  volume={338},
  number={1-2},
  pages={62--75},
  year={2022},
  publisher={Wiley Online Library}
}

@misc{weinhold2006epigenetics,
  title={Epigenetics: the science of change},
  author={Weinhold, Bob},
  year={2006},
  publisher={National Institute of Environmental Health Sciences}
}

@article{noller2024cell,
  title={Cell cycle expression heterogeneity predicts degree of differentiation},
  author={Noller, Kathleen and Cahan, Patrick},
  journal={Briefings in bioinformatics},
  volume={25},
  number={6},
  pages={bbae536},
  year={2024},
  publisher={Oxford University Press}
}

@article{eldar2010functional,
  title={Functional roles for noise in genetic circuits},
  author={Eldar, Avigdor and Elowitz, Michael B},
  journal={Nature},
  volume={467},
  number={7312},
  pages={167--173},
  year={2010},
  publisher={Nature Publishing Group UK London}
}

@article{azuara2006chromatin,
  title={Chromatin signatures of pluripotent cell lines},
  author={Azuara, V{\'e}ronique and Perry, Pascale and Sauer, Stephan and Spivakov, Mikhail and J{\o}rgensen, Helle F and John, Rosalind M and Gouti, Mina and Casanova, Miguel and Warnes, Gary and Merkenschlager, Matthias and others},
  journal={Nature cell biology},
  volume={8},
  number={5},
  pages={532--538},
  year={2006},
  publisher={Nature Publishing Group UK London}
}

@article{ng2008epigenetic,
  title={Epigenetic restriction of embryonic cell lineage fate by methylation of Elf5},
  author={Ng, Ray Kit and Dean, Wendy and Dawson, Claire and Lucifero, Diana and Madeja, Zofia and Reik, Wolf and Hemberger, Myriam},
  journal={Nature cell biology},
  volume={10},
  number={11},
  pages={1280--1290},
  year={2008},
  publisher={Nature Publishing Group UK London}
}

@article{sasai2013time,
  title={Time scales in epigenetic dynamics and phenotypic heterogeneity of embryonic stem cells},
  author={Sasai, Masaki and Kawabata, Yudai and Makishi, Koh and Itoh, Kazuhito and Terada, Tomoki P},
  journal={PLoS computational biology},
  volume={9},
  number={12},
  pages={e1003380},
  year={2013},
  publisher={Public Library of Science San Francisco, USA}
}

@article{grunstein1998yeast,
  title={Yeast heterochromatin: regulation of its assembly and inheritance by histones},
  author={Grunstein, Michael},
  journal={Cell},
  volume={93},
  number={3},
  pages={325--328},
  year={1998},
  publisher={Elsevier}
}

@article{zhang2019metabolic,
  title={Metabolic regulation of gene expression by histone lactylation},
  author={Zhang, Di and Tang, Zhanyun and Huang, He and Zhou, Guolin and Cui, Chang and Weng, Yejing and Liu, Wenchao and Kim, Sunjoo and Lee, Sangkyu and Perez-Neut, Mathew and others},
  journal={Nature},
  volume={574},
  number={7779},
  pages={575--580},
  year={2019},
  publisher={Nature Publishing Group UK London}
}

@article{hihara2012local,
  title={Local nucleosome dynamics facilitate chromatin accessibility in living mammalian cells},
  author={Hihara, Saera and Pack, Chan-Gi and Kaizu, Kazunari and Tani, Tomomi and Hanafusa, Tomo and Nozaki, Tadasu and Takemoto, Satoko and Yoshimi, Tomohiko and Yokota, Hideo and Imamoto, Naoko and others},
  journal={Cell reports},
  volume={2},
  number={6},
  pages={1645--1656},
  year={2012},
  publisher={Elsevier}
}

@article{kiefer2007epigenetics,
  title={Epigenetics in development},
  author={Kiefer, Julie C},
  journal={Developmental dynamics: an official publication of the American Association of Anatomists},
  volume={236},
  number={4},
  pages={1144--1156},
  year={2007},
  publisher={Wiley Online Library}
}

@article{gibney2010epigenetics,
  title={Epigenetics and gene expression},
  author={Gibney, ER and Nolan, CM},
  journal={Heredity},
  volume={105},
  number={1},
  pages={4--13},
  year={2010},
  publisher={Nature Publishing Group}
}

@article{reik2007stability,
  title={Stability and flexibility of epigenetic gene regulation in mammalian development},
  author={Reik, Wolf},
  journal={Nature},
  volume={447},
  number={7143},
  pages={425--432},
  year={2007},
  publisher={Nature Publishing Group UK London}
}

@article{cortini2016physics,
  title={The physics of epigenetics},
  author={Cortini, Ruggero and Barbi, Maria and Car{\'e}, Bertrand R and Lavelle, Christophe and Lesne, Annick and Mozziconacci, Julien and Victor, Jean-Marc},
  journal={Reviews of Modern Physics},
  volume={88},
  number={2},
  pages={025002},
  year={2016},
  publisher={APS}
}

@article{rue2015cell,
  title={Cell dynamics and gene expression control in tissue homeostasis and development},
  author={Ru{\'e}, Pau and Martinez Arias, Alfonso},
  journal={Molecular systems biology},
  volume={11},
  number={2},
  pages={792},
  year={2015}
}

@article{rando2009genome,
  title={Genome-wide views of chromatin structure},
  author={Rando, Oliver J and Chang, Howard Y},
  journal={Annual review of biochemistry},
  volume={78},
  number={1},
  pages={245--271},
  year={2009},
  publisher={Annual Reviews}
}
